\documentclass[prd,amsmath,showpacs,preprintnumbers,superscriptaddress,
               floatfix,twocolumn,letterpaper,
h-physrev4]{revtex4}

\usepackage{times}
\usepackage{graphicx}
\usepackage{slashed}
\usepackage{amssymb}
\usepackage{mathrsfs}
\newcommand{\nc}{\newcommand}
\nc{\lb}{\langle}
\nc{\rk}{\rangle}
\nc{\Blb}{\Big\langle}
\nc{\Brk}{\Big\rangle}
\nc{\rad}{\langle r^2\rangle}

\nc{\mi}{\!\!\mid\!\!}
\nc{\ra}{\rightarrow}
\nc{\Ra}{\Rightarrow}
\nc {\cd}{\partial}
\nc {\sla}{\slashed}
\nc{\ro}{\mathrm}
\nc{\ca}{\mathcal}
\nc{\sr}{\mathscr}
\nc{\bo}{\mathbf}
\nc{\Tr}{\ro{Tr}\,}
\nc{\Str}{\ro{Str}}
\nc{\realtrace}{\ro{Re\; Tr}}
\nc{\maxrealtrace}{\ro{max\, Re\; Tr}}
\nc{\ud}{\ro{d}}
\nc{\nn}{\nonumber}

\nc{\vk}{\vec{k}}
\nc{\vq}{\vec{q}}
\nc{\pb}{\bar{\psi}}
\nc{\p}{\psi}
\nc{\Pb}{\bar{\Psi}}
\nc{\vp}{\vec{\pi}}
\nc{\vap}{\varphi}
\nc{\vt}{\vec{\tau}}
\nc{\si}{\sigma}
\nc{\Si}{\Sigma}
\nc{\tSi}{\tilde{\Sigma}}
\nc{\g}{\gamma}
\nc{\G}{\Gamma}
\nc{\la}{\lambda}
\nc{\La}{\Lambda}
\nc{\ep}{\epsilon}
\nc{\de}{\delta}
\nc{\De}{\Delta}
\nc{\cL}{\ca{L}}
\nc{\cLe}{\ca{L}_{\ro{eff}}}
\nc {\ti}{\tilde}
\nc{\f}{\frac}
\nc{\da}{\dagger}
\nc{\SU}{\ro{SU}}
\nc{\om}{\omega}
\nc{\Om}{\Omega}

\nc{\darrow}{\stackrel{\leftrightarrow}{\cd}}
\nc{\darrows}{\stackrel{\leftrightarrow}{\sla{\cd}}}
\nc{\Darrows}{\stackrel{\leftrightarrow}{\sla{D}}}
\nc{\mr}{\stackrel{\circ}{m}_\rho}

\nc {\eqb}{\begin{equation}}
\nc {\eqe}{\end{equation}}
\nc {\eqab}{\begin{eqnarray}}
\nc {\eqae}{\end{eqnarray}}

\begin{document}

\title{Chiral extrapolations for nucleon electric charge radii}

\author{J. M. M. Hall} 
\affiliation{Special Research Centre for the Subatomic Structure of
  Matter (CSSM), School of Chemistry and Physics, University of Adelaide, 
  Adelaide, South Australia 5005, Australia}

\author{D. B. Leinweber}
\affiliation{Special Research Centre for the Subatomic Structure of
  Matter (CSSM), School of Chemistry and Physics, University of Adelaide, 
  Adelaide, South Australia 5005, Australia}

\author{R. D. Young} 
\affiliation{Special Research Centre for the Subatomic Structure of Matter 
  (CSSM), School of Chemistry and Physics, University of Adelaide, 
  Adelaide, South Australia 5005, Australia}
\affiliation{ARC Centre of Excellence for Particle Physics at the Terascale, 
School of Chemistry and Physics, University of Adelaide, 
  Adelaide, South Australia 5005, Australia}

\preprint{ADP-13-12/T832}

\begin{abstract}

Lattice simulations 
 for the electromagnetic form factors 
of the nucleon yield insights into the internal structure of hadrons. 
The logarithmic divergence of the charge radius in the chiral limit 
poses an interesting challenge in achieving reliable predictions from 
finite-volume lattice simulations. 
Recent results near the physical pion mass 
($m_\pi \sim 180\,\,{\rm MeV}$) 
are examined in order to confront the issue of how the 
chiral regime is approached. 
The electric charge radius of the nucleon isovector 
presents a forum for achieving consistent finite-volume corrections. 
Newly developed techniques within the framework of chiral effective 
field theory ($\chi$EFT) are used 
to achieve a robust extrapolation of the 
electric charge radius to the physical pion mass, and to infinite volume. 
The chiral extrapolations exhibit considerable finite-volume dependence;  
  lattice box sizes of $L\gtrsim 7\,{\rm fm}$
 are required in order to achieve a direct lattice simulation 
result within $2\%$ of the infinite-volume 
 value at the physical point. Predictions of the volume dependence are 
provided to guide the interpretation of future lattice results. 




\end{abstract}


\maketitle

\section{Introduction}
\label{sect:intro}

Much experimental progress has been made \cite{Gao:2003ag,HydeWright:2004gh,Arrington:2006zm,Perdrisat:2006hj,Arrington:2011kb} in 
examining the internal structure of hadrons, 
particularly with regard to the internal distribution of electric and magnetic 
charge due to quarks. 
Current understanding of the 
internal charge distribution, characterized by 
the elastic form factors, is also fortified by developments in 
supercomputing power and lattice QCD techniques.  Lattice QCD has seen 
significant advances 
in simulating  electromagnetic form factors, 
and is now able to probe the chiral 
regime \cite{Yamazaki:2009zq,Syritsyn:2009mx,Bratt:2010jn,Alexandrou:2011db}.

Recent results from the QCDSF Collaboration, using pion masses of order  
$\sim \! 180\,\,{\rm MeV}$ \cite{Collins:2011mk}, provide a new opportunity 
for exploring the utility 
 of chiral effective field theory 
($\chi$EFT)-based 
techniques  
in performing an extrapolation to the physical point. 
Additional care must be taken in 
 handling finite-volume effects relating to
 the electric charge radius 
\cite{Tiburzi:2007ep,Hu:2007eb,Jiang:2008ja,Hall:2012yx}.
In order to address this issue, a variety of  \textit{Ans\"{a}tze} 
for the $Q^2$ behaviour of the form factor  
are examined in order to construct a finite-volume analogue. 
The finite-volume corrections are applied directly to the electric 
form factors, and 
the electric charge radii are then calculated at infinite volume. 
By combining these methods with new techniques within the framework of 
$\chi$EFT, a robust extrapolation to the physical regime 
is performed herein.

In performing a chiral extrapolation, 
one should ideally 
use lattice simulation results 
 that lie within the chiral power-counting regime (PCR) of chiral 
perturbation theory in order to 
avoid a regularization scheme-dependent result. 
The PCR is defined by the range of quark (or pion) masses at which 
a $\chi$EFT calculation is independent of the regularization scheme, 
and typically lies in a pion-mass range of $\lesssim 200$ MeV 
\cite{Leinweber:2003dg,Leinweber:2005cm,Leinweber:2005xz}. 
Within the PCR, the chiral expansion of an observable is a controlled 
expansion, 
and the result is insensitive to treatments of higher-order terms, such as 
the resummation of the chiral series. 
Since lattice QCD results usually extend outside the PCR, one is restricted 
by the available 
data when performing an extrapolation. 
An important application of finite-range regularization (FRR) is the ability 
to extrapolate using lattice QCD results that 
extend beyond the PCR. 
One method for achieving this involves
identifying a preferred regularization scale 
and 
an upper bound of the pion mass directly from the lattice QCD 
results, as demonstrated in Refs.~\cite{Hall:2010ai,Hall:2011en}. 
In a previous investigation,  
 a successful extrapolation of the magnetic moment of the nucleon 
to the physical point was achieved using these techniques \cite{Hall:2012pk}. 
This analysis similarly provides a prediction of the pion-mass 
dependence of the electric charge radius of the nucleon  
for a range of lattice volumes. 

\begin{figure}[tp]
\includegraphics[height=0.950\hsize,angle=90]{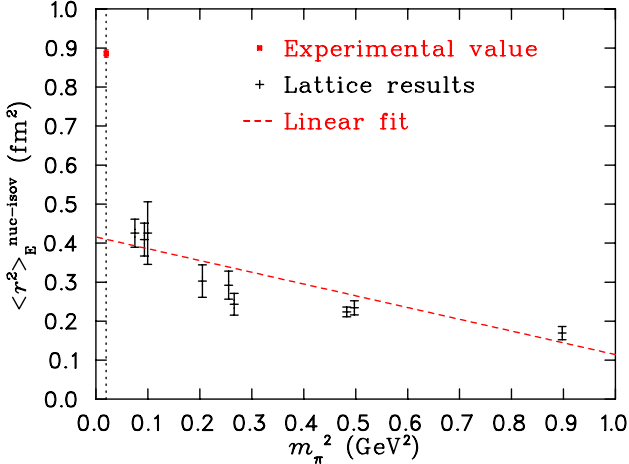}
\vspace{-11pt}
\caption{\footnotesize{(color online). Lattice QCD results for $\rad_E$ 
from QCDSF \cite{Collins:2011mk}, using the Ansatz from Eq.~(\ref{eqn:quad}), 
and the experimental value as marked \cite{Mohr:2012tt,Beringer:1900zz}. The lattice results satisfy: $L>1.5$ fm and $m_\pi L>3$. A na\"{i}ve linear
 trend line is also included, which does not reach the experimental value. The physical point is shown with a vertical dotted line.}}
\label{fig:data}
\end{figure}

The lattice QCD results from the QCDSF Collaboration \cite{Collins:2011mk} 
used in this analysis are displayed 
in Fig.~\ref{fig:data}. 
The simulation used a two-flavor $\ca{O}(a)$-improved Wilson quark action,    
and the isovector nucleon $(p - n)$ was calculated to avoid 
the computational cost of disconnected loops that occur in full QCD. 
Only the simulation results that satisfy the criteria: 
$L>1.5$ fm and $m_\pi L > 3$, are shown. Of the nine points that satisfy 
these criteria,  the lattice size varies from $1.7$ to $2.9$ fm. 
The QCDSF results are displayed using
 a Sommer scale parameter of  $r_0 = 0.475$ fm, based on 
results from Ref.~\cite{Aubin:2004wf}. 
Without consideration of chiral loop 
contributions, it is clear 
that there would be a factor of two 
discrepancy between the lattice QCD simulations 
and the experimental value \cite{Mohr:2012tt,Beringer:1900zz}
as shown by a linear trend line. 

\vspace{-3mm}
\section{Chiral effective field theory}
\label{sect:eft}
\vspace{-1mm}
\subsection{Electromagnetic form factors}
\vspace{-1mm}
It is common to 
define the Sachs electromagnetic form factors $G_{E,M}$, which parametrize 
the matrix element 
 for the quark current $J_\mu$. In the 
heavy-baryon limit, 
this can be written as  
\begin{align}
\lb B(p')| J_\mu|B(p)\rk &= \bar{u}^{s'}(p')\Big\{\ro{v}_\mu\, G_E(Q^2)
\nonumber\\
&+\f{i\ep_{\mu\nu\rho\si}\ro{v}^\rho\,
 S_\ro{v}^\sigma\, q^\nu}{m_B}\,G_M(Q^2)\Big\}u^s(p),
\label{eqn:matelem}
\end{align}
where $Q^2$ is defined as positive momentum transfer $Q^2 = -q^2 = -(p'-p)^2$.
Lattice QCD results are often constructed from an alternative representation, 
using the form factors $F_1$ and $F_2$,  
 the Dirac and Pauli form factors, respectively.
The Sachs form factors are simply linear combinations of $F_1$ and $F_2$, 
\begin{align}
G_E(Q^2) &= F_1(Q^2) - \f{Q^2}{4m_B^2}F_2(Q^2)\,,\\
G_M(Q^2) &= F_1(Q^2) + F_2(Q^2)\,.
\end{align}
In the heavy-baryon formulation of the quark current matrix element 
shown in Eq.~(\ref{eqn:matelem}), 
the spin operator  
$S^\mu_\ro{v} = -\f{1}{4}\g_5[\g^\mu,\g^\nu]\ro{v}_\nu$ 
is required. It 
has the useful properties in that its commutation and anticommutation
rules depend only on the four-velocity of the baryon $\ro{v}_\mu$ 
\cite{Jenkins:1990jv,Jenkins:1991ts}.
The momentum-dependent 
electric form factor $G_E(Q^2)$ 
allows a charge radius 
to be defined in the usual manner, 
\eqb
\label{eqn:raddefn}
\rad_E = \lim_{Q^2\rightarrow 0}-6\frac{\cd G_E(Q^2)}{\cd Q^2}.
\eqe

For the leading-order contributions to the electric form factor, the 
following first-order interaction 
Lagrangian from heavy-baryon chiral perturbation theory ($\chi$PT) is used 
\cite{Jenkins:1991ne,Jenkins:1990jv,Jenkins:1991ts,Labrenz:1996jy,WalkerLoud:2004hf,Wang:2007iw}, 
\begin{align}
\label{eqn:lag}
\cL^{(1)}_{\chi PT}  &= 2D\,\Tr[\bar{B}_\ro{v}S_\ro{v}^\mu\{A_\mu,B_\ro{v}\}\,]
+2F\,\Tr[\bar{B}_\ro{v}S_\ro{v}^\mu[A_\mu,B_\ro{v}]\,] \nn \\
&+\ca{C}\,(\bar{T}_\ro{v}^\mu A_\mu B_\ro{v} +
 \bar{B}_\ro{v}A_\mu  T_\ro{v}^\mu).
\end{align}
The pseudo-Goldstone fields $\xi(x)$ 
are encoded in the adjoint representation of 
$\SU(3)_L\otimes\SU(3)_R$, forming an axial vector combination, denoted
$A_\mu$, 
\begin{align}
\xi &\equiv \ro{exp}\left\{{\f{i}{f_\pi}\tau^a \pi^a}\right\},\\
A_\mu &= \f{1}{2}(\xi\,\cd_\mu\,\xi^\da - \xi^\da\,\cd_\mu\,\xi).
\end{align}
The values for the $D$, $F$ and $\ca{C}$ couplings in 
the interaction Lagrangian are related through 
 $\SU(6)$ flavor-symmetry 
\cite{Jenkins:1991ts,Lebed:1994ga}, $F = \f{2}{3}D$
and $\ca{C} = -2D$. Phenomenological values of the constants  
 $D = 0.76$ and $f_\pi = 92.4$ MeV are used. 

\subsection{Finite-range regularization}
\label{subsect:frr}

In FRR $\chi$EFT, 
a regulator function $u(k\,;\La)$, with characteristic momentum scale $\La$, 
is introduced in the numerators of the loop integrals. The regulators 
should be chosen such that they satisfy   
$u|_{k=0} = 1$ and $u|_{k\rightarrow\infty} = 0$.
The result of an FRR calculation is independent of the choice of 
$u(k\,;\La)$ if the lattice simulation points are 
constrained entirely within the PCR. 
In this investigation, a dipole form is chosen,  which takes 
the following form, 
\eqb
u(k\,;\La) = {\left(1 + \f{k^2}{\La^2}\right)}^{-2}.
\eqe
While conventional $\chi$PT fails outside the PCR, FRR $\chi$EFT remains 
effective, as the regulator takes on an additional role in modelling 
the effect of higher-order terms in the expansion. 
Analyses have been undertaken previously for a range of possible forms of 
regulator function 
\cite{Leinweber:2005xz,Hall:2010ai}. 

\subsection{Loop integrals and definitions}
\label{subsect:loop}

\begin{figure}
\centering
\includegraphics[height=80pt]{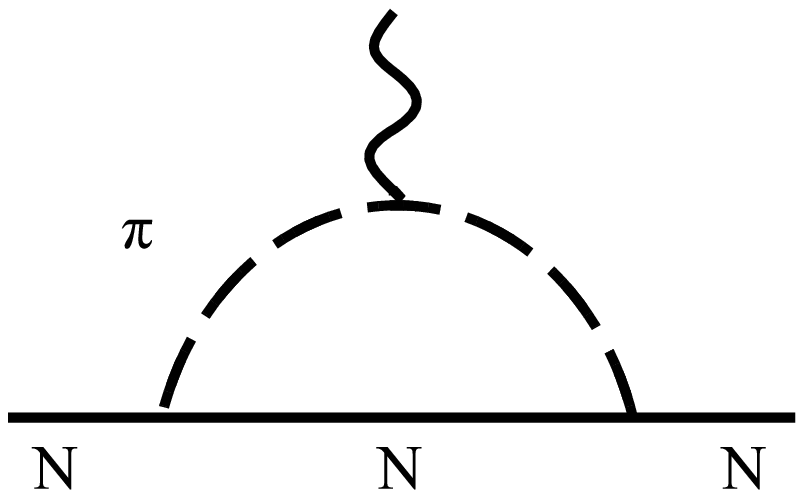}
\caption{\footnotesize{The pion loop contributions to the electric charge radius of a nucleon. All charge conserving pion-nucleon transitions are implicit.}}
\label{fig:emSEa}
\vspace{2mm}
\centering
\includegraphics[height=80pt,angle=0]{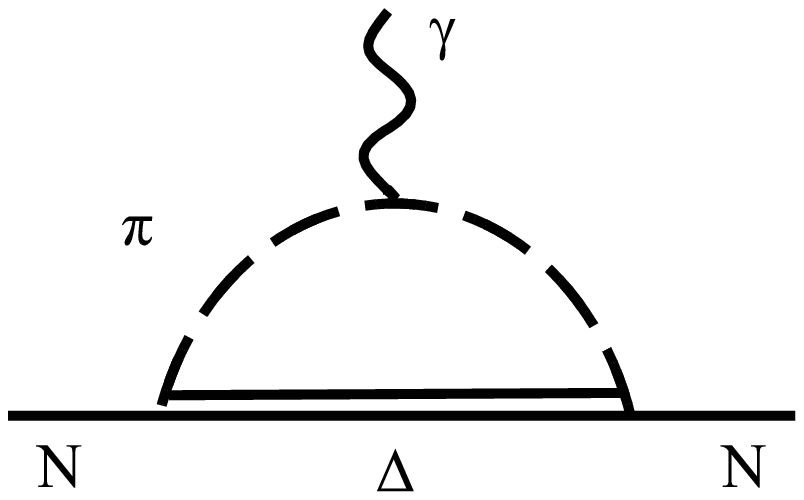}
\caption{\footnotesize{The pion loop contribution to the electric charge radius
 of a nucleon, allowing transitions to the nearby and strongly-coupled $\Delta$ baryons.}}
\label{fig:emSEb}
\vspace{4mm}
\centering
\includegraphics[height=80pt,angle=0]{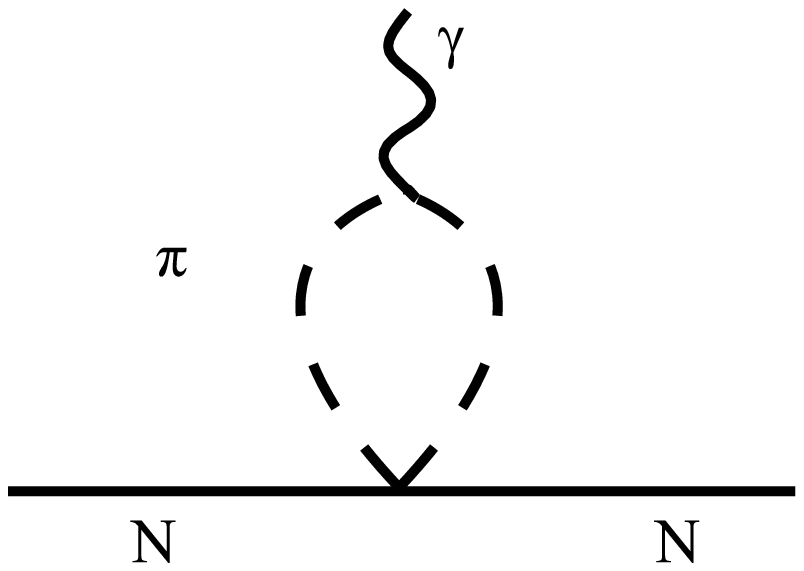}
\caption{\footnotesize{The tadpole contribution at $\ca{O}(m_q)$ to the 
electric charge radius of a nucleon.}}
\label{fig:emSEc}
\end{figure}

The leading-order loop integral contributions to the electric form factor 
correspond to the diagrams in Figs.~\ref{fig:emSEa} through \ref{fig:emSEc}.  
The electric charge radius itself is also renormalized by contributions from 
loop integrals, obtained from $\chi$EFT.  
The loop integrals 
can be simplified to a
convenient form by taking the  heavy-baryon limit, and
performing the pole integrations for $k_0$,   
\begin{align}
\label{eqn:SiL}
\ca{T}^E_{N}(Q^2) &= 
-\frac{\chi^E_{N}}{5\pi}\!\int\!\!\mathrm{d}^3 k 
\frac{(k^2-\vk\cdot\vq)\,u(\vk\,;\La)\,u(\vk-\vq\,;\La)}
{\omega_{\vk}\omega_{\vk-\vq}(\omega_{\vk}
+\omega_{\vk-\vq})},\\
\ca{T}^E_{\De}(Q^2) &= 
-\frac{\chi^E_{\De}}{5\pi}\!\int\!\!\mathrm{d}^3 k 
\frac{(k^2-\vk\cdot\vq)\,u(\vk\,;\La)\,u(\vk-\vq\,;\La)}
{(\omega_{\vk} + \Delta)(\omega_{\vk-\vq} + \Delta)(\omega_{\vk}
+\omega_{\vk-\vq})},\\
\ca{T}^E_{\ro{tad}}(Q^2) &= -\frac{\chi^E_{t}}{\pi}\!\int\!\!\mathrm{d}^3 
k \frac{u^2(\vk\,;\La)}
{\omega_{\vk}+\omega_{\vk-\vq}},
\label{eqn:SiT}
\end{align}
where $\om_{\vec{k}} = \sqrt{{\vec{k}}^2 + m^2_\pi}$ 
and $\De$ is the baryon mass splitting. 

The coefficients $\chi_N^E$, $\chi_\De^E$ and $\chi_t^E$, for both proton ($p$) 
and neutron ($n$), 
are related to the constants $D$, $F$, $\ca{C}$ and $f_\pi$ 
from the chiral Lagrangian in Eq.~(\ref{eqn:lag}),   
\begin{align}
\chi_N^{E,p} &= -\f{5}{16\pi^2f_\pi^2}(D+F)^2 = -\chi_N^{E,n},\\
\chi_\De^{E,p} &= +\f{5}{16\pi^2f_\pi^2}\f{4\ca{C}^2}{9} = -\chi_\De^{E,n},\\
\chi_t^{E,p} &= -\f{1}{16\pi^2f_\pi^2} = -\chi_t^{E,n}.
\end{align}

\subsection{Finite-volume corrections}

Finite-volume corrections 
cannot be applied 
directly to the charge radius itself \cite{Hall:2012yx}. 
Instead, the electric form factors $G_E(Q^2)$ are corrected to 
infinite-volume. 
To obtain the integrals $T^E$ that 
contribute to the electric charge radius, 
one takes the derivative of $\ca{T}^E$ 
with respect to momentum transfer $\vec{q}^2$, as 
$\vec{q}^2 \rightarrow 0$,  
\eqb
T^E = \lim_{\vec{q}^2\rightarrow 0}-6\f{\cd \ca{T}^E(\vec{q}^2)}{\cd \vec{q}^2}, 
\label{eqn:siSi}
\eqe 
which is equivalent to the derivative in Eq.~(\ref{eqn:raddefn}) 
in the Breit frame, defined by $q=(0,\vec{q})$.

The finite-volume corrections 
to the electric form factors are achieved by subtracting
the electric charge symmetry-preserving finite-volume correction, defined as  
\eqb
\label{eqn:fvc}
 \Delta_L(Q^2,0) = 
\delta_L\!\left[\ca{T}^E(Q^2)\right]-\delta_L\!\left[\ca{T}^E(0)\right]. 
\eqe
The functional $\delta_L$ is defined through the convention 
\cite{Armour:2005mk}: 
\eqb
\delta_L[\ca{T}^E(Q^2)] = 
\chi\left[\f{{(2\pi)}^3}{L_x L_y L_z}\sum_{k_x,k_y,k_z} -%
\int\!\mathrm{d}^3k\right]\ca{I}^E(Q^2),
\eqe
for an integrand $\ca{I}^E$.
The
second term of Eq.~(\ref{eqn:fvc}) ensures that both infinite- and 
finite-volume electric form 
factors are correctly normalized, 
i.e. 
$G_E(0) = 1$. 
This normalization procedure exploits the lattice Ward Identity 
 that ensures charge conservation 
is satisfied in a finite volume. It has been shown previously 
that this is realised in practice; numerically 
and through $\chi$EFT analyses \cite{Hall:2012yx}. 
Thus, the infinite-volume electric form factor can be calculated 
using the equation: 
\eqb
G_E^\infty(Q^2) = G_E^L(Q^2) - \Delta_L(Q^2,0).
\eqe
The infinite-volume charge radius $\rad_E^\infty$ can be recovered 
from the form factor by choosing an \textit{Ansatz} for the extrapolation in $Q^2$, 
analogous to the procedure typically performed at finite volume. 

In applying FRR to the finite-volume corrections, the value of 
$\Delta_L(Q^2,0)$ stabilises as $\Lambda$ becomes large. 
Applying the same technique as in Ref~\cite{Hall:2010ai}, 
the asymptotic result of $\Delta_L(Q^2,0)$ is achieved numerically 
by evaluating it with a dipole regulator, 
using a relatively large value of $\La' = 2.0$ GeV. 
This method  is similar 
 to the algebraic approach outlined in
Ref.~\cite{Beane:2004tw}, and has been successfully demonstrated 
in previous studies \cite{AliKhan:2003cu}.

\subsection{Renormalization}

The procedure for the renormalization of the low-energy coefficients 
of the chiral expansion in FRR $\chi$EFT will now be outlined. A thorough 
discussion can be found in Ref.~\cite{Hall:2010ai}.

Each loop integral contributing to the electric charge radius
 may be expanded out as an analytic polynomial plus a nonanalytic term, 
\begin{align}
T^E_N(m_\pi^2\,;\La) &= b_0^{\La,N} + \chi^E_N \log\f{m_\pi}{\mu} + 
b_2^{\La,N}m_\pi^2 + 
\ca{O}(m_\pi^4),\\
T^E_\De(m_\pi^2\,;\La) &= b_0^{\La,\De}  + b_2^{\La,\De}m_\pi^2 
+  \frac{\chi^E_\De}{2\De^2}\,m_\pi^2\log\f{m_\pi}{\mu} + \ca{O}(m_\pi^4),\\
T^E_\ro{tad}(m_\pi^2\,;\La) &= 
b_0^{\La,t} + \chi^E_t \log\f{m_\pi}{\mu} + b_2^{\La,t}m_\pi^2 + 
\ca{O}(m_\pi^4),
\end{align}
where $\mu$ is a mass scale associated with the chiral logarithm. 
Once the lattice results have been converted into infinite-volume charge radii, 
the chiral behaviour of the electric charge radius 
can 
be written in terms of an ordered expansion in pion-mass squared, 
through use of the Gell-Mann$-$Oakes$-$Renner
Relation, $m_q \propto m_\pi^2$ \cite{GellMann:1968rz}, 
\begin{align}
\label{eqn:chiralexp}
\rad_E^\infty &= \{a_0^\La + a_2^\La m_\pi^2\}+ T^E_{N}(m_\pi^2\,;\La) 
+ T^E_\De(m_\pi^2\,;\La) \nonumber \\
&+ T^E_{\ro{tad}}(m_\pi^2\,;\La) + \ca{O}(m_\pi^4).
\end{align}
This expansion contains an analytic polynomial in $m_\pi^2$ plus 
the leading-order chiral loop integrals, 
from which nonanalytic behaviour arises.
The scale-dependent coefficients 
$a_i^\La$ are the residual series coefficients, which
correspond to direct quark-mass insertions in the full
Lagrangian. 
Upon renormalization of the divergent loop
integrals, these will correspond with low-energy coefficients of $\chi$EFT 
\cite{Young:2002ib}. 

In order to obtain the renormalized chiral coefficients, $c_i$, 
one must add the $b_i^\La$ 
terms from each of the loop integrals 
to the residual series coefficients 
$a_i^\La$, 
\begin{align}
c_0 = a_0^\La + b_0^{\La,N} + b_0^{\La,\De} + b_0^{\La,t},\\
c_2 = a_2^\La + b_2^{\La,N} + b_2^{\La,\De} + b_2^{\La,t}. 
\end{align}
The resultant coefficients, $c_0$ and $c_2$, 
are the renormalized low-energy
 coefficients of the chiral expansion at the scale, $\mu$.
By evaluating the loop integrals, the renormalized 
chiral expansion can also be written in terms of a polynomial in $m_\pi^2$ 
and the nonanalytic terms,  
\begin{align}
\rad_E^\infty &= c_0^{(\mu)} + (\chi_N^E +
\chi_t^E)\log\f{m_\pi}{\mu} + c_2 m_\pi^2 \nn\\
&+  \frac{\chi_\De^E}{2\De^2}\,m_\pi^2\log\f{m_\pi}{\mu} +\ca{O}(m_\pi^4)\,, 
\label{eqn:logexpn}
\end{align}
reproducing $\chi$PT in the PCR. 
Since 
the chiral 
expansion of Eq.~(\ref{eqn:logexpn}) contains a logarithm, 
the value of $c_0$ can only be extracted relative to the mass scale, $\mu$, 
which is chosen to be $1$ GeV in this case. 

To achieve a chiral extrapolation, 
it is convenient to subtract the $b_0^\La$ coefficients 
 from the respective loop integrals, thus automating the 
renormalization procedure to chiral order $\ca{O}(1)$, 
\begin{align}
\tilde{T}_N^E &= T_N^E - b_0^{\La,N},\\
\tilde{T}_\De^E &= T_\De^E - b_0^{\La,\De},\\
\tilde{T}_\ro{tad}^E &= T_\ro{tad}^E - b_0^{\La,t}.
\end{align}
This removes the dependence on the 
regularization scale $\La$ 
in the leading low-energy coefficient. 
Thus, the chiral formula used for fitting lattice 
QCD results takes the form: 
\eqb
\label{eqn:chiralexprenorm}
\rad_E^\infty = \{c_0^{(\mu)} + a_2^\La m_\pi^2\} + \tilde{T}^E_{N} + 
\tilde{T}^E_\De +\tilde{T}^E_{\ro{tad}} + \ca{O}(m_\pi^4).
\eqe
To ascertain the presence of an optimal regularization scale $\La^{\ro{scale}}$, 
the renormalization flow of the leading low-energy 
coefficient $c_0^{(\mu)}$ will be considered 
in Sec.~\ref{subsect:curves}, using the prescription detailed in 
Refs.~\cite{Hall:2010ai,Hall:2011en,Hall:2012pk}. 

\section{Results}
\label{sect:results}

\subsection{$Q^2$ extrapolation}
\label{subsect:Qsqextrap}
\vspace{-2mm}

\begin{figure}[tp]
\includegraphics[height=0.950\hsize,angle=90]{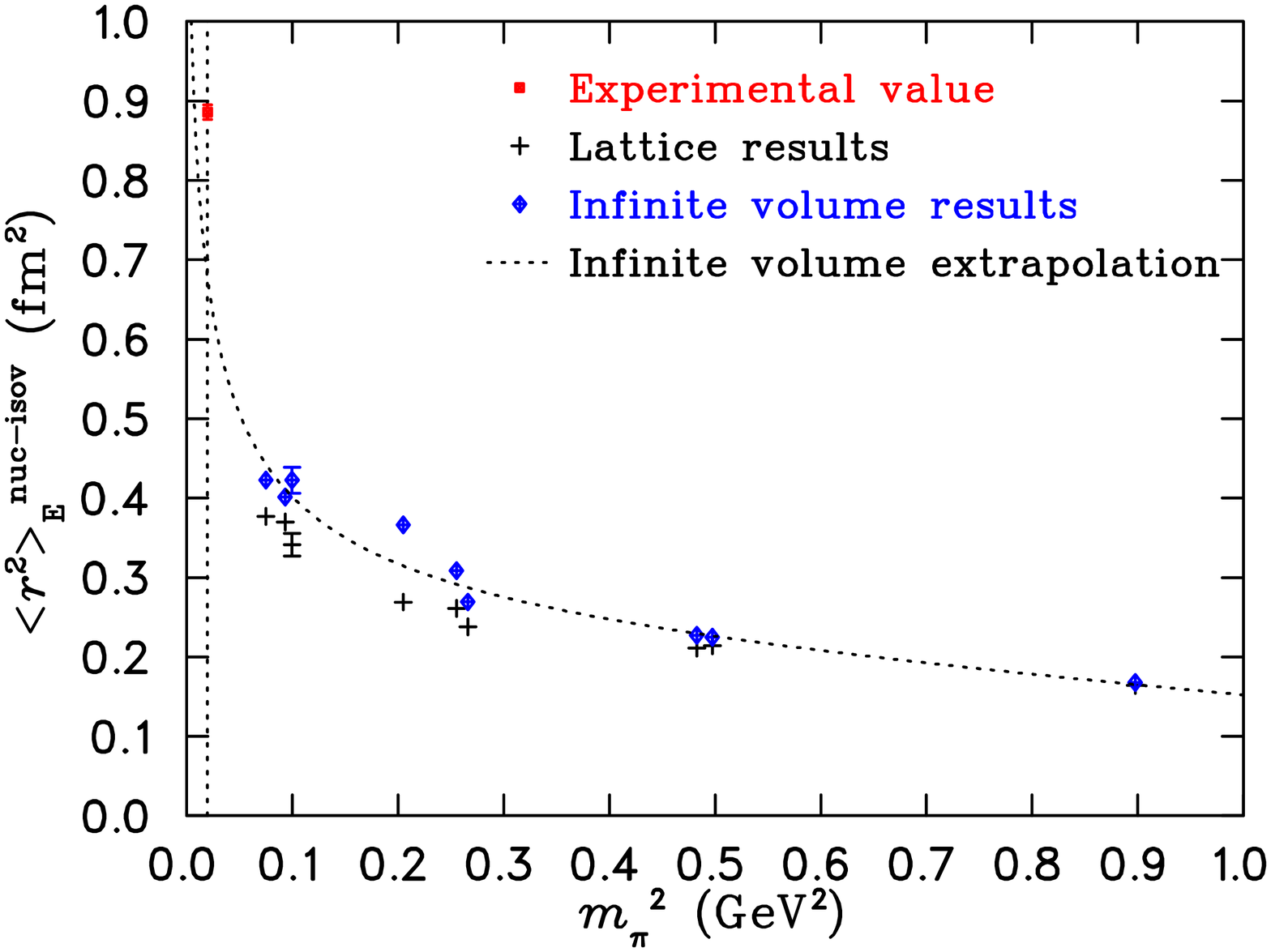}
\vspace{-11pt}
\caption{\footnotesize{(color online). Infinite-volume chiral extrapolation of $\rad_E$,  using the dipole $Q^2$ extrapolation \textit{Ansatz} from Eq.~(\ref{eqn:dip}).  The infinite-volume corrected lattice points are also shown.}}
\label{fig:dipnormal}
\vspace{6mm}
\includegraphics[height=0.950\hsize,angle=90]{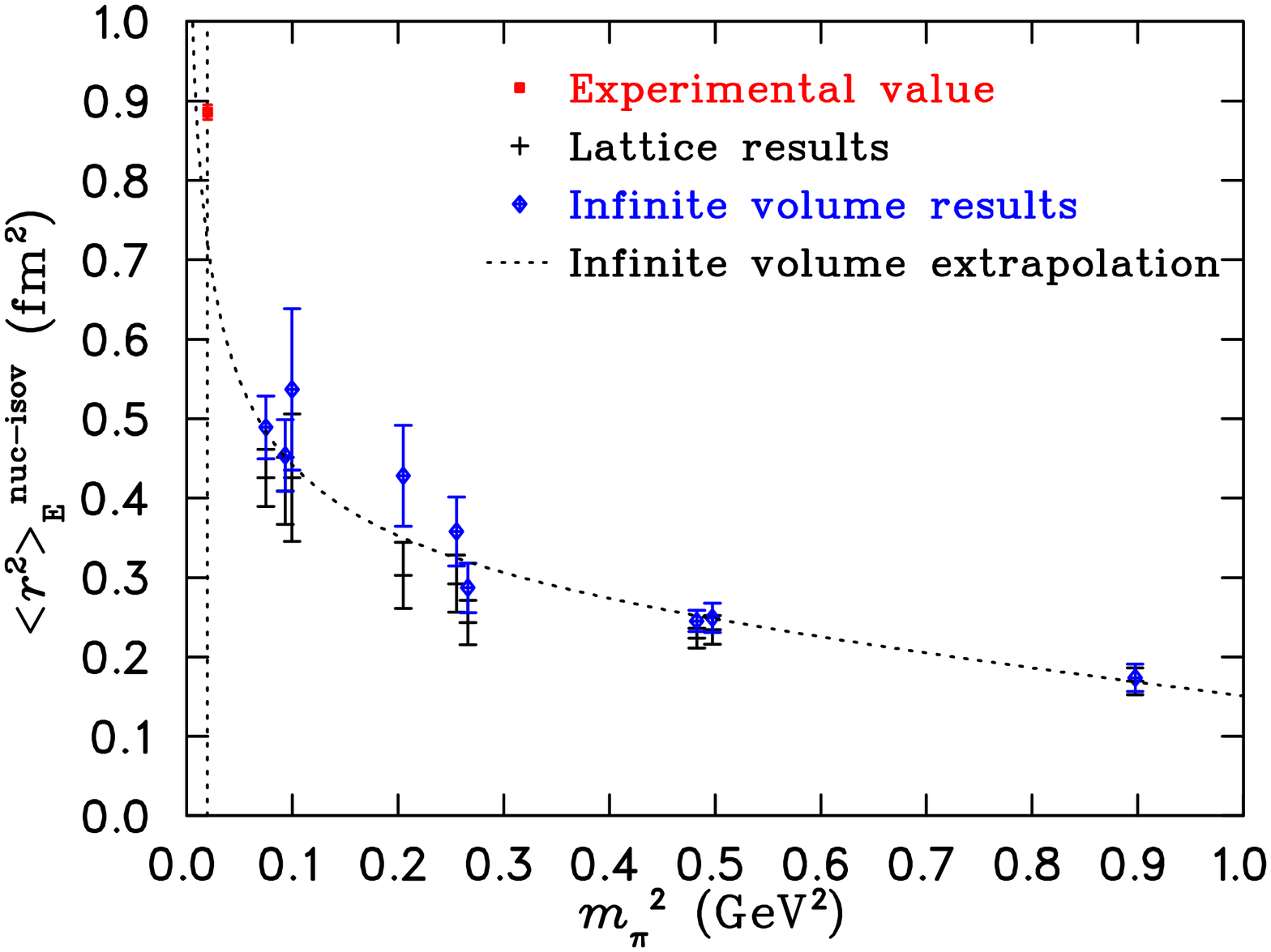}
\vspace{-11pt}
\caption{\footnotesize{(color online). Infinite-volume chiral extrapolation of $\rad_E$, using the variant $Q^2$ extrapolation \textit{Ansatz} from Eq.~(\ref{eqn:quad}).
}}
\label{fig:quad}
\end{figure}

In extracting an electric charge radius 
from typical lattice QCD results on periodic volumes,  
one must choose an \textit{Ansatz} to model the finite-volume corrected 
$Q^2$ behaviour of the electric 
form factor. A common choice is the dipole form, defined by 
\eqb
\label{eqn:dip}
G_E(Q^2) = \f{G_E(0)}{(1+Q^2/\La_D^2)^2},
\eqe
where the dipole mass $\La_D$ is a free parameter, related to the electric 
charge radius 
by  $\La_D^2 = 12/\rad_E$. This \textit{Ansatz} tightly constrains the $Q^2$ dependence 
and leads to small errors in 
 the radius $\rad_E$ compared with other \textit{Ans\"{a}tze}. 
These dipole-constrained radii are shown in Fig.~\ref{fig:dipnormal}.  

A modification may be made 
to account for higher order terms in $Q^2$. An inverse quadratic with two 
fit parameters, as inspired by Kelly \cite{Kelly:2004hm}, may be chosen.  
This form is used in the analysis 
 by the QCDSF Collaboration \cite{Collins:2011mk},   
\eqb
\label{eqn:quad}
G_E(Q^2) = \f{G_E(0)}{1 + \alpha Q^2 + \beta Q^4}. 
\eqe
The charge radius is obtained through  $\rad_E = 6\,\alpha$. 
This \textit{Ansatz} 
was originally chosen for modelling the 
large $Q^2$ behaviour of $F_1$ \cite{Kelly:2004hm,Collins:2011mk}. 
 However, it is of greater interest here 
to examine and compare the \emph{small} $Q^2$ behaviour of this \textit{Ansatz} with that 
of the dipole. Furthermore, this will provide a guide to the expected 
variation in $\rad_E$ due to the choice 
of \textit{Ansatz}. 

\begin{figure}[tp]
\includegraphics[height=0.950\hsize,angle=90]{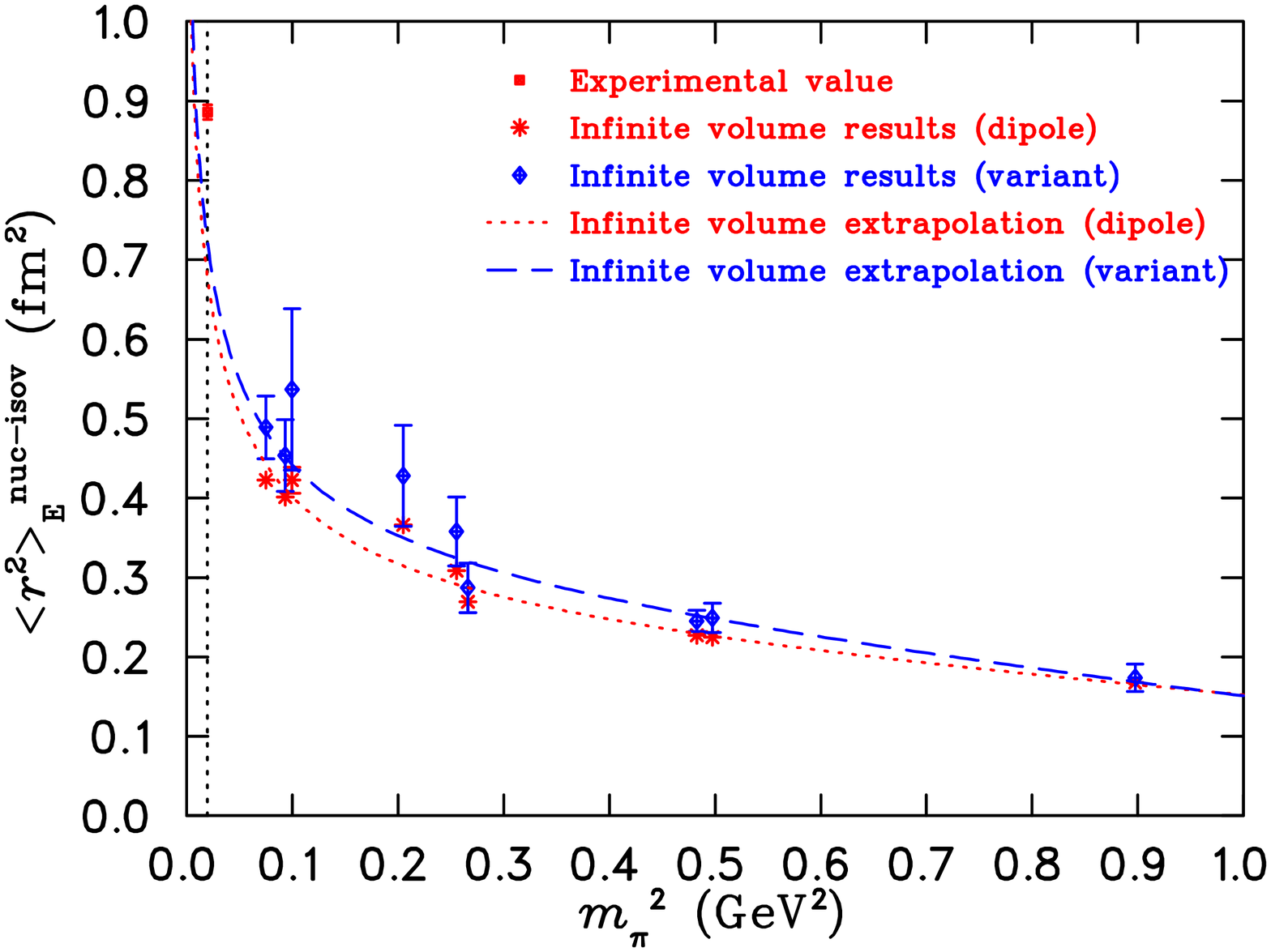}
\vspace{-11pt}
\caption{\footnotesize{(color online). A comparison of the infinite-volume chiral extrapolations of $\rad_E$ using the dipole $Q^2$ extrapolation \textit{Ansatz} from Eq.~(\ref{eqn:dip}), and the variant $Q^2$ extrapolation \textit{Ansatz} from Eq.~(\ref{eqn:quad}).}}
\label{fig:qdcomp}
\vspace{6mm}
\includegraphics[height=0.950\hsize,angle=90]{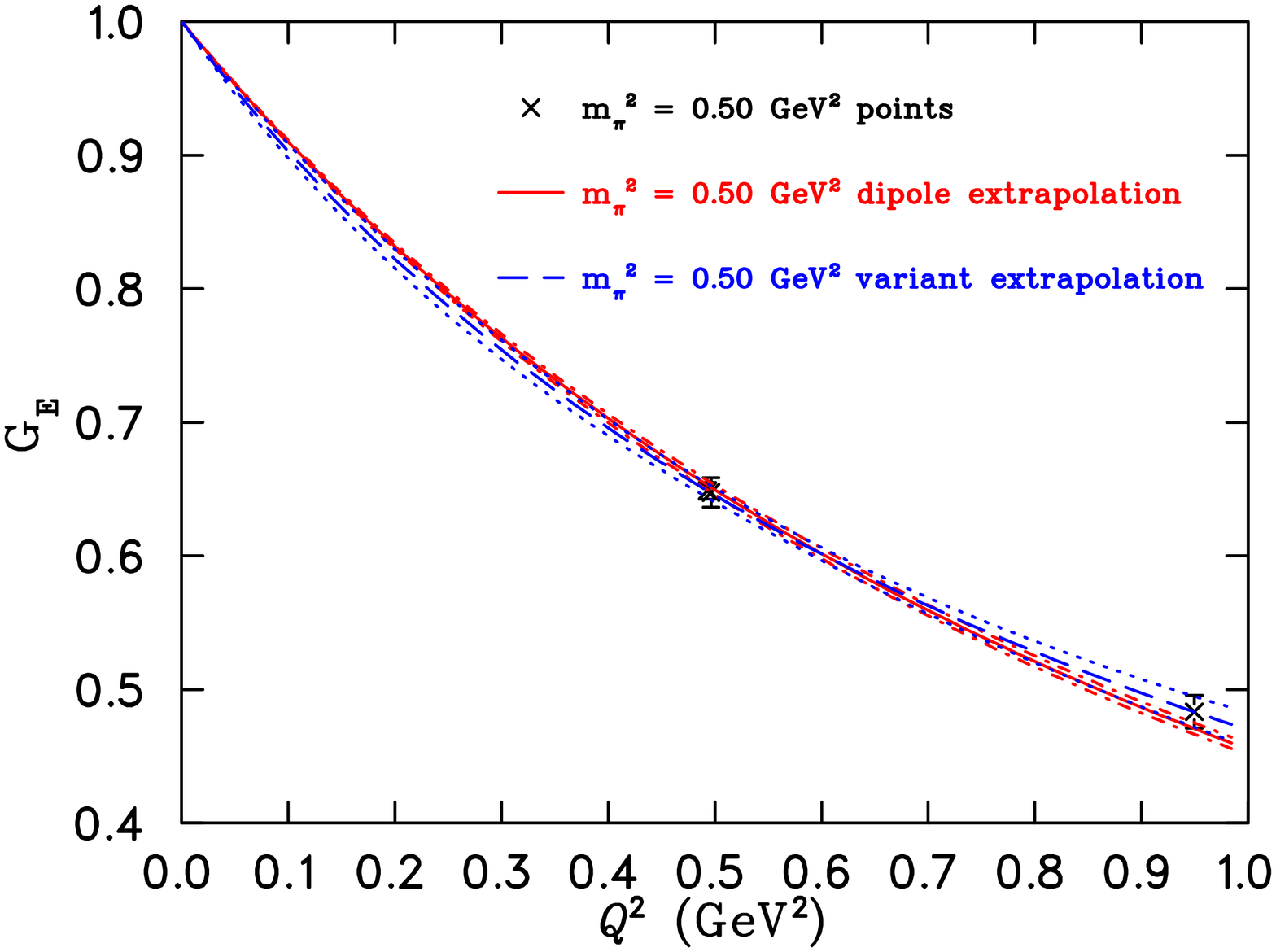}
\vspace{-11pt}
\caption{\footnotesize{(color online). A comparison of the $Q^2$ extrapolation of the electric form factor $G_E$, using the normal dipole \textit{Ansatz} from Eq.~(\ref{eqn:dip}), and the variant \textit{Ansatz}, defined in Eq.~(\ref{eqn:quad}). 
The smallest three values of $Q^2$ are used (the smallest two being almost coincident). The fits are shown for $m_\pi^2 = 0.50$ GeV$^2$. Error bands are shown with dotted lines.}}
\label{fig:Qsqcomp}
\end{figure}

A demonstration of an infinite-volume chiral extrapolation using each \textit{Ansatz} 
is shown in Figs.~\ref{fig:dipnormal} and \ref{fig:quad}. 
In each case, the smallest three $Q^2$ values available are considered 
in fitting the \textit{Ansatz} parameters. For illustrative purposes, FRR is performed 
with a dipole regulator with $\Lambda = 1.0$ GeV. 
A direct 
comparison of the finite-volume-corrected lattice values of $\rad_E$, using 
the dipole \textit{Ansatz} from Eq.~(\ref{eqn:dip}), and the variant \textit{Ansatz} 
from Eq.~(\ref{eqn:quad}), is shown in Fig.~\ref{fig:qdcomp}. 

In Fig.~\ref{fig:dipnormal}, the estimate of the uncertainty in $\rad_E$ 
is much smaller than for the other \textit{Ansatz}, raising concerns of an 
unaccounted for systematic uncertainty. The electric charge radii obtained 
using the variant 
\textit{Ansatz} from Eq.~(\ref{eqn:quad}), as shown in Fig.~\ref{fig:quad}, appear to be 
the more cautious, in that the error bar encompasses a range of variation
 from the  
choice of \textit{Ansatz}. This can be seen most clearly in Fig.~\ref{fig:qdcomp}.  

In order to assess the low $Q^2$ behaviour of the variant \textit{Ansatz} in   
Eq.~(\ref{eqn:quad}), 
a comparison of the $Q^2$ extrapolation using this \textit{Ansatz} 
 is shown in Fig.~\ref{fig:Qsqcomp} at the 
point:  $m_\pi^2 = 0.50$ GeV$^2$. Both $Q^2$ 
extrapolations are plotted on the same axes. 
The lightest 
three values of $Q^2$ are used in constraining the parameters. The merit 
of the extra fit parameter in Eq.~(\ref{eqn:quad}) is evident.  

\vspace{-3mm}
\subsection{Renormalization flow analysis}
\label{subsect:curves}
\vspace{-2mm}

The QCDSF results  
for the electric charge radius, displayed in Fig.~\ref{fig:data}, 
include a linear extrapolation, 
which does not take into account 
the nonanalytic behaviour of the 
chiral loop integrals, nor the finite-volume corrections. 
Neglecting these important 
effects \cite{Leinweber:1992hj}, it is not surprising that the 
 linear trend line does not approach the experimental value of the 
electric charge radius at the physical pion mass.
Since these lattice QCD results extend outside the PCR, the result of 
an extrapolation will be regularization scale dependent. 
However, the scale dependence
 may be 
constrained using a procedure \cite{Hall:2010ai,Hall:2011en,Hall:2012pk} 
that obtains an optimal regularization 
scale, and an estimate of its uncertainty, as constrained 
by the lattice results. 

\begin{figure}[tp]
\includegraphics[height=0.950\hsize,angle=90]{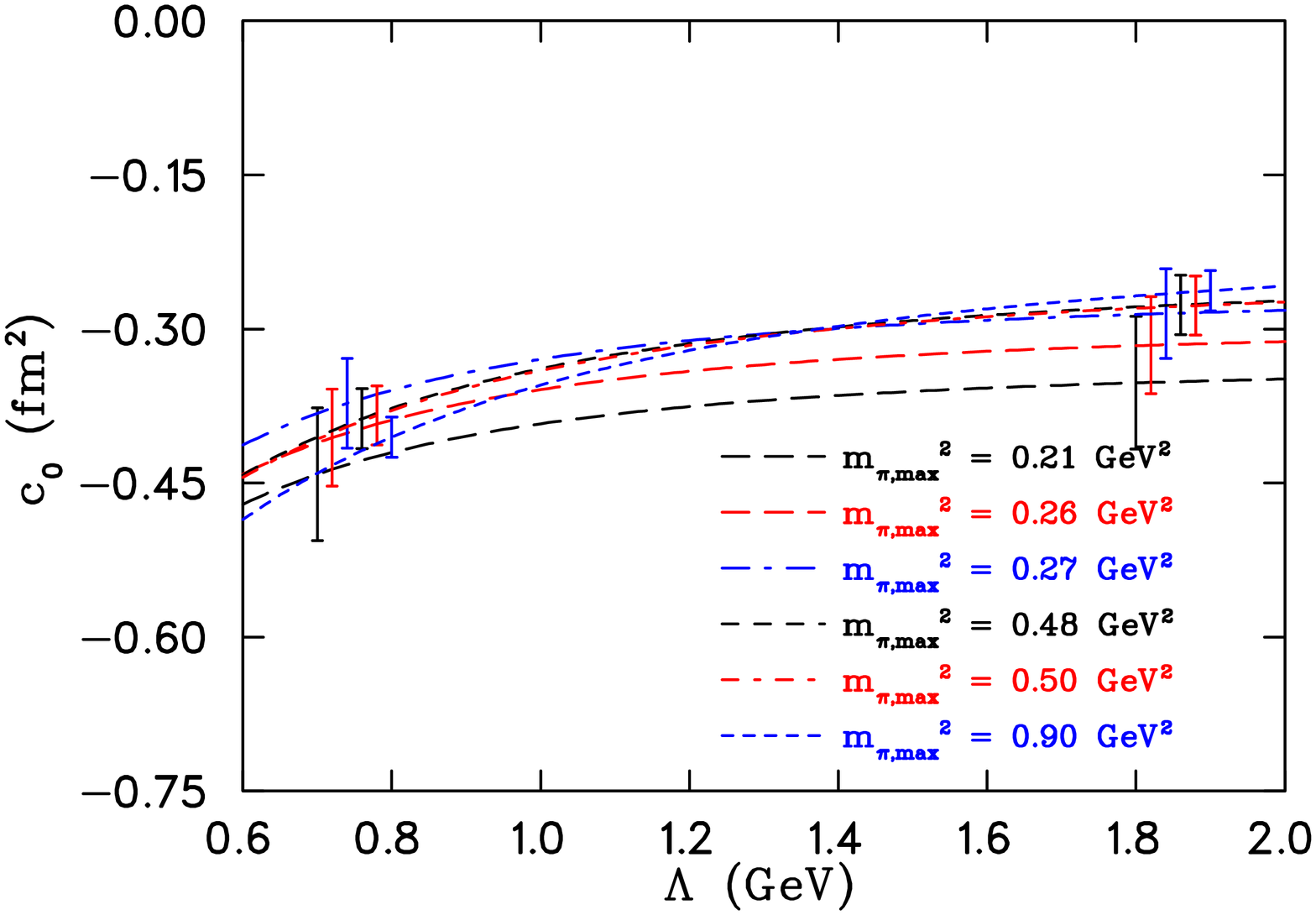}
\vspace{-11pt}
\caption{\footnotesize{(color online). The renormalization flow of
  $c_0^{(\mu)}$, obtained using a dipole regulator, and based on  
 QCDSF simulation results. $c_0^{(\mu)}$ 
is calculated relative to the mass scale, 
$\mu = 1$ GeV. For each curve, two arbitrary values of $\La$ are chosen to indicate the general size of the error bars.}}
\label{fig:c0}
\vspace{6mm}
\includegraphics[height=0.950\hsize,angle=90]{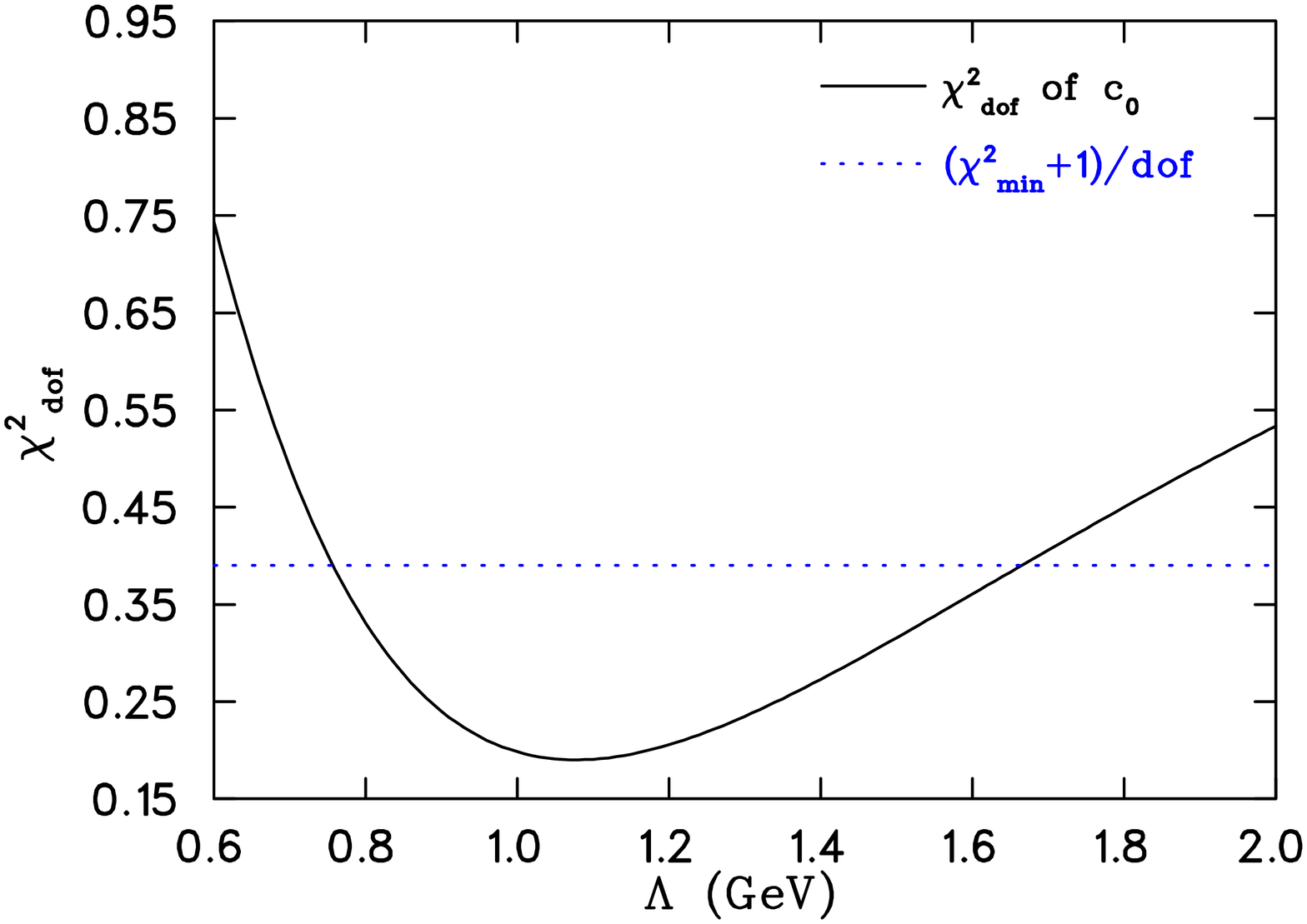}
\vspace{-11pt}
\caption{\footnotesize{(color online). A $\chi^2_{dof}$ analysis for 
 the renormalization flow of
  $c_0^{(\mu)}$, obtained using a dipole regulator, and based on  
 QCDSF simulation results. The dotted line illustrates the upper limit:  
$(\chi^2_{\ro{min}}+1)/dof$. 
}}
\label{fig:c0chisqdof}
\end{figure}

In order to obtain an optimal regularization scale, 
the low-energy coefficient, $c_0^{(\mu)}$ 
from Eq.~(\ref{eqn:chiralexprenorm}), will be calculated across a range 
of values of the regularization scale, $\La$. 
Multiple renormalization flow curves may be obtained by constraining the 
fit window by a maximum value, $m_{\pi,\ro{max}}^2$, and sequentially adding  
 points to extend further outside the PCR. 
The renormalization flow curves for a dipole regulator 
are plotted on the same set of axes 
in Fig.~\ref{fig:c0}. 
Within the PCR, $c_0$ will be insensitive to the value of $\Lambda$, and 
appear as a horizontal line in Fig.~\ref{fig:c0}. 
In contrast, variation of $c_0$ with respect to $\Lambda$ becomes 
larger as one moves further from the PCR. The correct value of $c_0$, 
and thus the optimal value for $\Lambda$, is identified 
by the intersections of the curves, where their deviation is minimal 
\cite{Hall:2010ai}. 

Unlike the results from the 
analysis of the nucleon mass \cite{Hall:2010ai} 
or magnetic moment \cite{Hall:2012pk}, the regularization scale-dependence 
is relatively weak for $\Lambda > 0.6$ GeV. 
There is no distinct intersection point 
in the renormalization flow curves.
 This lack of sensitivity to the regularization
 scale is a consequence of the logarithm 
in the chiral expansion of Eq.~(\ref{eqn:logexpn}), 
which is slowly-varying with respect to $\La$. 

An optimal regularization scale for the dipole regulator 
can be obtained 
using a $\chi^2_{dof}$ analysis, 
 taking the degrees of freedom to be the 
curves of $c_0$ corresponding to different 
values of $m_{\pi,\ro{max}}^2$. 
For the six different values of $m_{\pi,\ro{max}}^2$ considered 
in Fig.~\ref{fig:c0}, each curve is described by $c_0^i(\Lambda)$, 
where $i$ takes values $1$ through $6$. $\de c_0^i(\La)$ denotes 
the uncertainty in $c^i_0$ obtained when fitting the lattice results. 
The $\chi^2_{dof}$ for each value of $\Lambda$ is expressed as 
\begin{align}
\chi^2_{dof}(\Lambda) = 
\f{1}{n-1} \sum_{i=1}^{n} \f{{(c_0^i(\La) - \bar{c}_0(\La))}^2}
{{(\de c^i_0(\La))}^2},\\
\bar{c}_0(\La) = \f{\sum_{i=1}^{n}c_0^i(\La)/{{(\de c_0^i(\La))}^2}}
{\sum_{j=1}^{n} 1 / {(\de c_0^j(\La))}^2},
\label{eqn:av}
\end{align}
with the statistically weighted average $\bar{c}_0(\La)$ given by 
Eq.~(\ref{eqn:av}). The $\chi^2_{dof}$ is illustrated in 
Fig.~\ref{fig:c0chisqdof}. 
The value of the optimal scale, obtained using a dipole regulator, is  
$\La^{\ro{scale}}_{\ro{dip}} = 1.08^{+0.58}_{-0.32}$ GeV, which is  
 consistent with the optimal regularization scale 
values obtained 
for the nucleon mass 
using a dipole regulator \cite{Hall:2010ai}. 
The value 
is also consistent 
with the result 
obtained for the nucleon magnetic moment, based on these 
QCDSF simulations \cite{Hall:2012pk}. 
This provides evidence that, 
for a given functional form of the regulator, the optimal 
regularization scale 
may be associated with an intrinsic scale, characterizing the finite size of 
the nucleon. 

\vspace{-2mm}
\subsection{Chiral extrapolations}
\label{subsect:extrap}
\vspace{-1mm}

The identification of an optimal regularization scale 
 allows an accurate chiral extrapolation to be performed. 
Furthermore, a range of box sizes may be considered, thus providing 
 an estimate of the finite-volume effects.
In order to determine the most suitable number of  points to be used 
for fitting the lattice results, the method described in 
Refs.~\cite{Hall:2011en,Hall:2012pk} is used. 

In extrapolating the electric charge radius, 
the statistical uncertainty 
comprises contributions from the fit coefficients.   
In the case of the systematic uncertainty, the axial coupling  
and the pion decay constant are assumed to be sufficiently 
well-determined experimentally. Thus, 
the dominant contribution to the systematic uncertainty 
in the extrapolation is associated with the optimal regularization scale. 

A second source of systematic uncertainty is due to the 
choice of the regulator functional form, which is combined in quadrature, 
\begin{equation}
\Big({\delta\rad_E^{\ro{sys}}}\Big)^2 = \Big({\delta\rad_E^{\La}}\Big)^2
+\Big({\delta\rad_E^{\ro{reg}}}\Big)^2.
\end{equation}
 $\delta\rad_E^{\ro{reg}}$ 
is obtained by comparing the result of a dipole regulator 
to that of using a sharp cutoff, which has an intrinsic scale 
of $\La^{\ro{scale}}_{\ro{sc}} = 0.51^{+0.17}_{-0.10}$ GeV, determined using the 
same methods described for the dipole regulator. 
The systematic uncertainty is taken as 
half the difference between the central values of each case. Though the 
sharp cutoff regulator does not provide higher-order nonanalytic contributions 
in the chiral expansion \cite{Djukanovic:2004px}  
and is less physical than the dipole regulator, a comparison between 
the two regulators provides the most cautious evaluation of the dependence 
of the result on the functional form of the regulator. 

\begin{figure}[tp]
\includegraphics[height=0.95\hsize,angle=90]{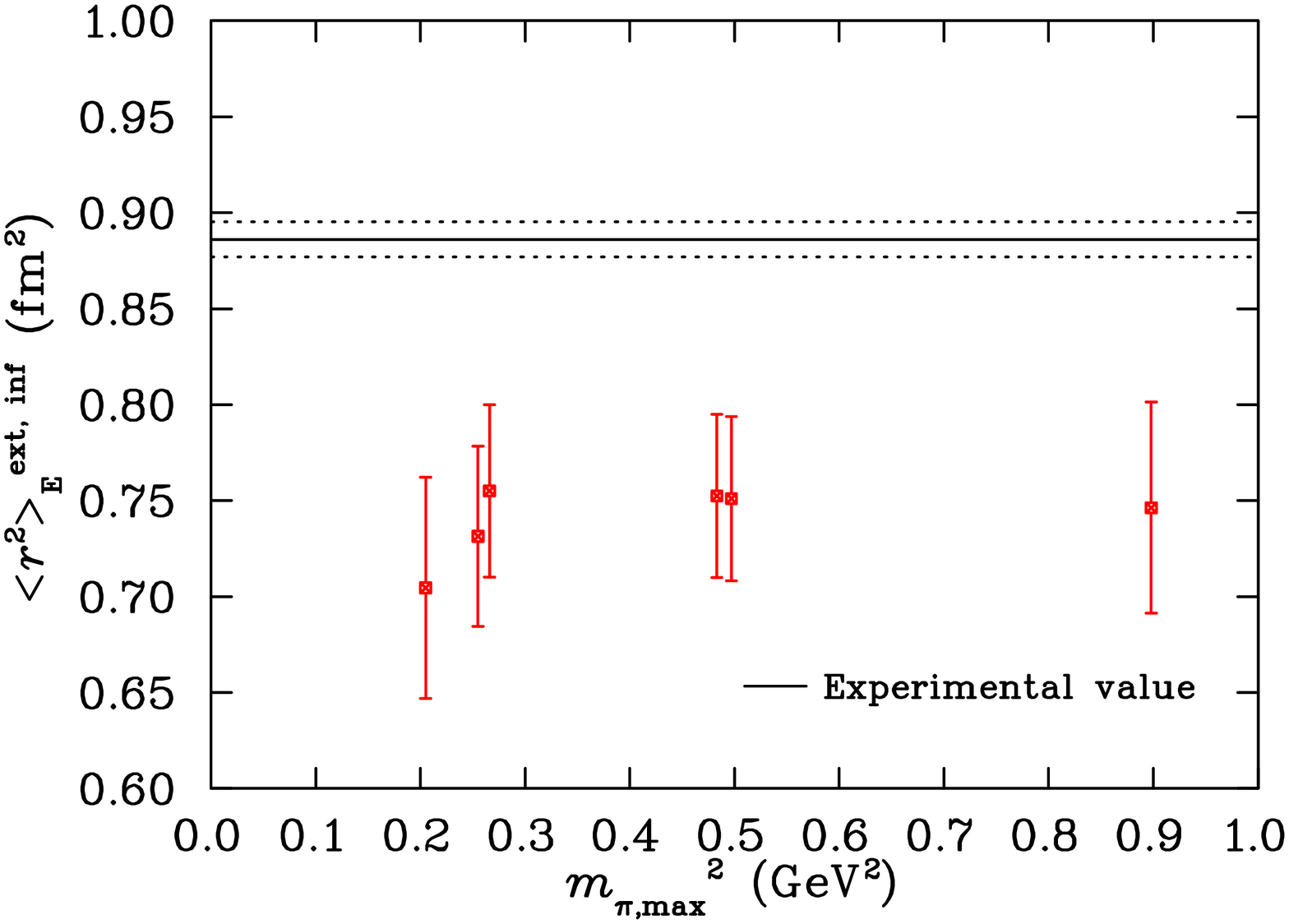}
\vspace{-11pt}
\caption{\footnotesize{(color online). Behaviour of the extrapolation of $\rad_E$ to the physical point, vs.\ $m_{\pi,\ro{max}}^2$. The 
value of $\La^{\ro{scale}}$ is used, 
as obtained from the $\chi^2_{dof}$ analysis. The error bars include the statistical and systematic uncertainties 
added in quadrature. 
}}
\label{fig:revsmpisq}
\vspace{3mm}
\includegraphics[height=0.95\hsize,angle=90]{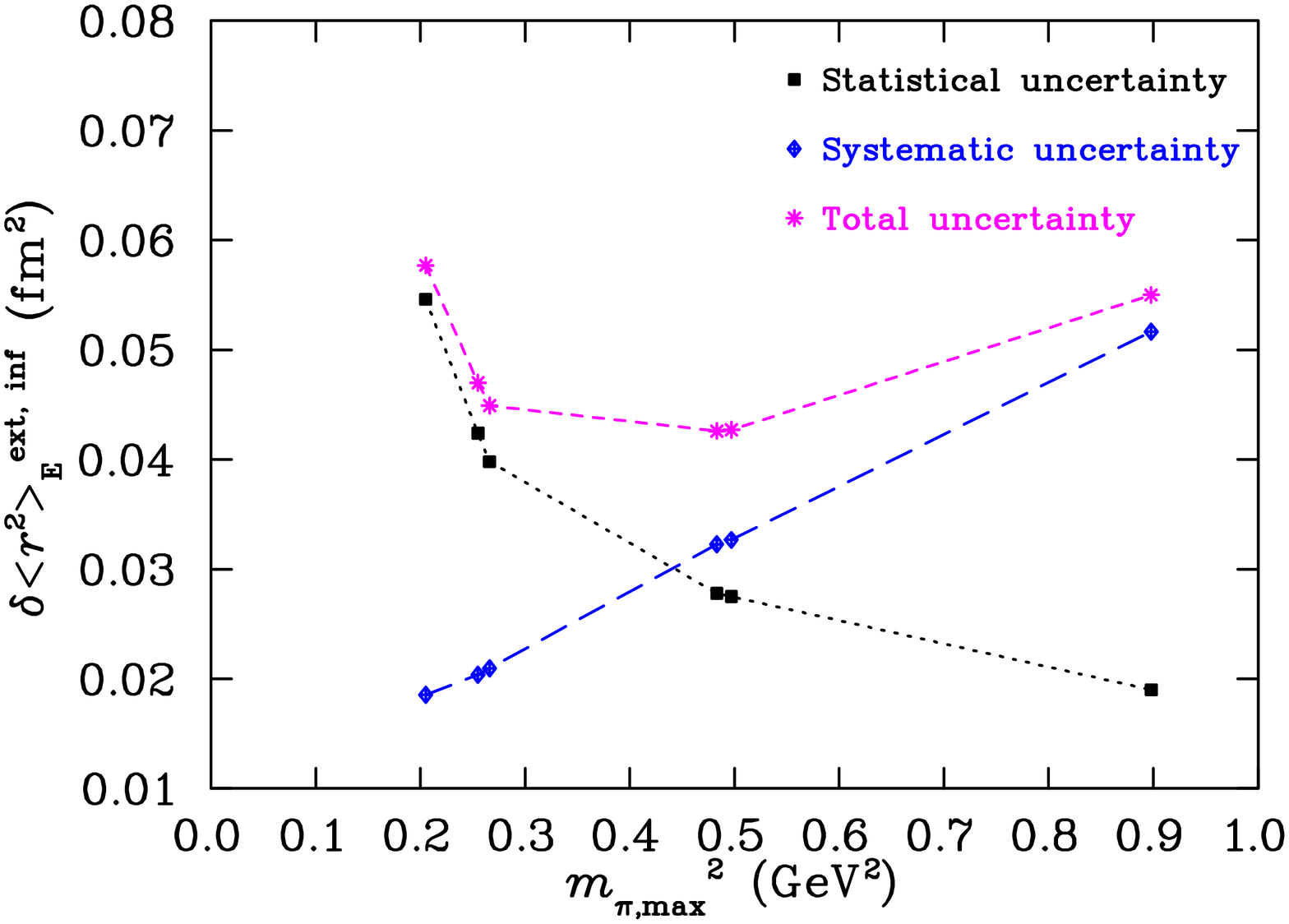}
\vspace{-11pt}
\caption{\footnotesize{(color online). Magnitude of the statistical, systematic and total error bars in the extrapolation of $\rad_E$ to the physical point, vs.\ $m_{\pi,\ro{max}}^2$. In each case of regulator, the value of $\La^{\ro{scale}}$ is used, as obtained from the corresponding $\chi^2_{dof}$ analysis.  At a maximum pion mass of $\hat{m}_{\pi,\ro{max}}^2 = 0.48$ GeV$^2$, the best compromise between statistical and systematic uncertainty is achieved. }}
\label{fig:revsmpisqss}
\end{figure}

The value of the extrapolation of $\rad_E$ to the physical point is shown in 
Fig.~\ref{fig:revsmpisq} for different values of 
$m_{\pi,\ro{max}}^2$. Statistical and systematic errors have been added 
in quadrature. 
Fig.~\ref{fig:revsmpisqss} shows the magnitude of the statistical and 
systematic error bars separately, in addition to the total uncertainty. 
These plots allow the identification of the optimal number of lattice results 
to be used for an extrapolation, 
which is signified by the best compromise between the 
statistical and the systematic uncertainties. 
Figure~\ref{fig:revsmpisqss} indicates that, in this case,
the lightest seven lattice points should be used, 
corresponding to a value of $m_{\pi,\ro{max}}^2 \simeq 0.48$ GeV$^2$.
Table \ref{table:ss} summarizes the 
breakdown of each error bar into its source components.

Note that there is a discrepancy between the experimental value 
and the extrapolation results. This could be 
a consequence of excited state contamination in the 
lattice calculation of the three-point function; the use of only two flavours, 
and/or neglecting $\ca{O}(a)$ effects.

\begin{table*}[tp]
  \caption{\footnotesize{Results for the isovector nucleon  electric charge radius, extrapolated to the physical point using different values of $m_{\pi,\ro{max}}^2$, as illustrated in Fig.~\ref{fig:revsmpisq}. The uncertainty in $\rad_E\,(m_{\pi,\ro{phys}}^2)$ is provided in the following order: the statistical uncertainty, 
the uncertainty due to $\La^{\ro{scale}}$, the uncertainty due to the change in regulator functional form, and the total uncertainty, respectively. The value of $\La^{\ro{scale}}$ is calculated for each choice of regulator functional form. 
}}
\vspace{-6pt}
  \newcommand\T{\rule{0pt}{2.8ex}}
  \newcommand\B{\rule[-1.4ex]{0pt}{0pt}}
  \begin{center}
    \begin{tabular}{llllll}
      \hline
      \hline
      \T\B            
      $m_{\pi,\ro{max}}^2$(GeV$^2$)  &  \, $\rad_E(m_{\pi,\ro{phys}}^2)$ (fm$^2$)&  \, $\delta\rad_E^{\ro{stat}}$ &  \quad $\delta\rad_E^{\La}$ &  \quad $\delta\rad_E^{\ro{reg}}$ &  \quad $\delta\rad_E^{\ro{tot}}$ \\
      \hline     
      $0.205$   &\T\,\, $0.705$ & \,\,$0.055$ & \quad$0.017$ & \quad$0.006$ & \quad$0.058$\\
      $0.255$   &\T\,\, $0.731$ & \,\,$0.042$ & \quad$0.019$ & \quad$0.007$ & \quad$0.047$\\
      $0.266$   &\T\,\, $0.755$ & \,\,$0.040$ & \quad$0.020$ & \quad$0.007$ & \quad$0.045$\\
      $0.483$   &\T\,\, $0.753$ & \,\,$0.028$ & \quad$0.031$ & \quad$0.008$ & \quad$0.043$\\ 
      $0.497$   &\T\,\, $0.751$ & \,\,$0.028$ & \quad$0.032$ & \quad$0.008$ & \quad$0.043$\\
      $0.898$   &\T\,\, $0.746$ & \,\,$0.019$ & \quad$0.051$ & \quad$0.007$ & \quad$0.055$\\
      \hline
    \end{tabular}    
  \end{center}
  \label{table:ss}
\end{table*}

\vspace{-3mm}
\subsection{Finite-volume effects in future lattice simulations}
\vspace{-1mm}

\begin{figure*}
\includegraphics[height=0.50\hsize,angle=90]{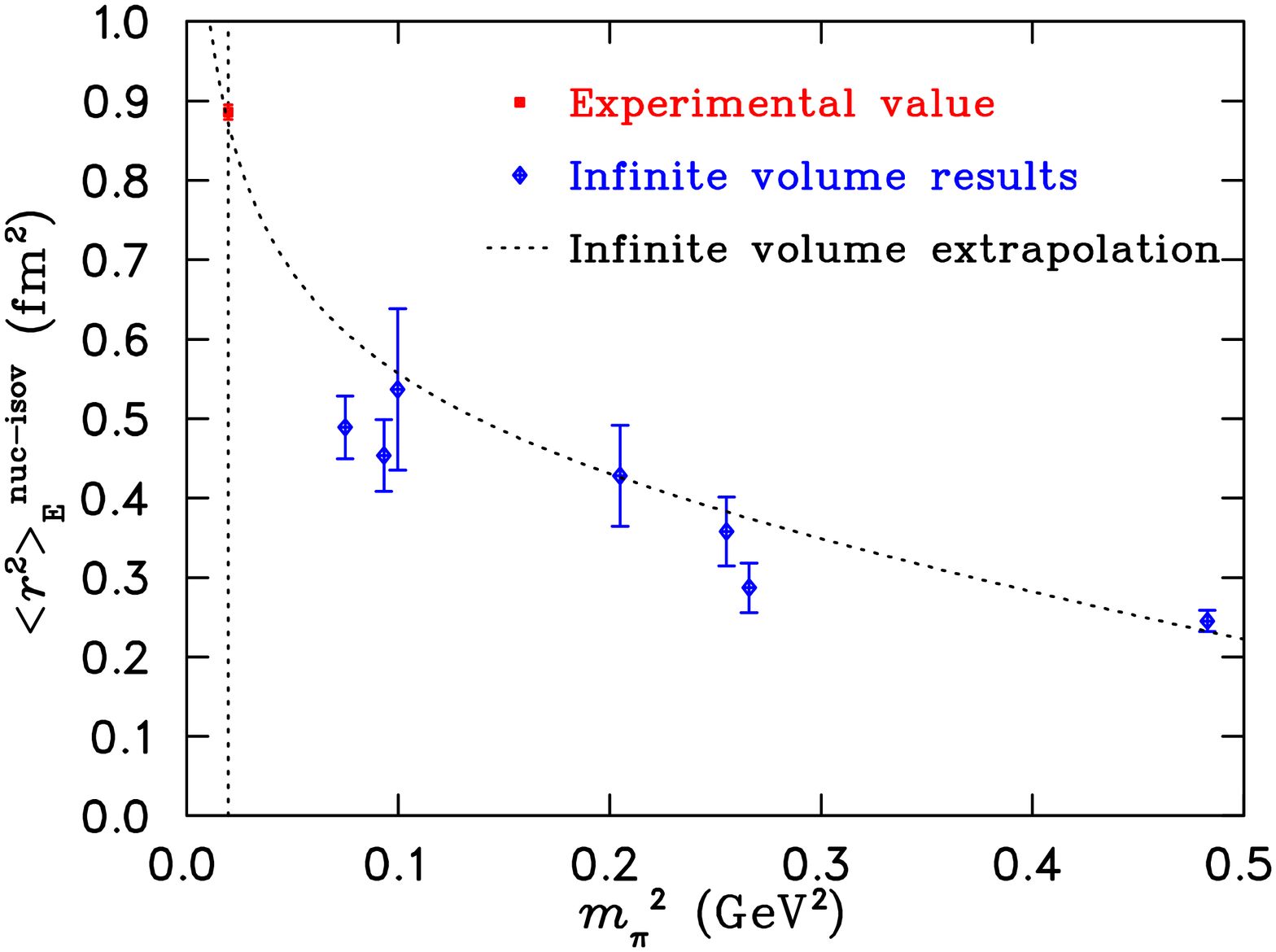}
\vspace{-11pt}
\caption{\footnotesize{(color online). Extrapolation of $\rad_E$  
at infinite volume. The experimental value has been included in the fit, in preparation for making future finite-volume corrections.    
}}
\label{fig:extrapinf}
\vspace{3mm}
\includegraphics[height=0.50\hsize,angle=90]{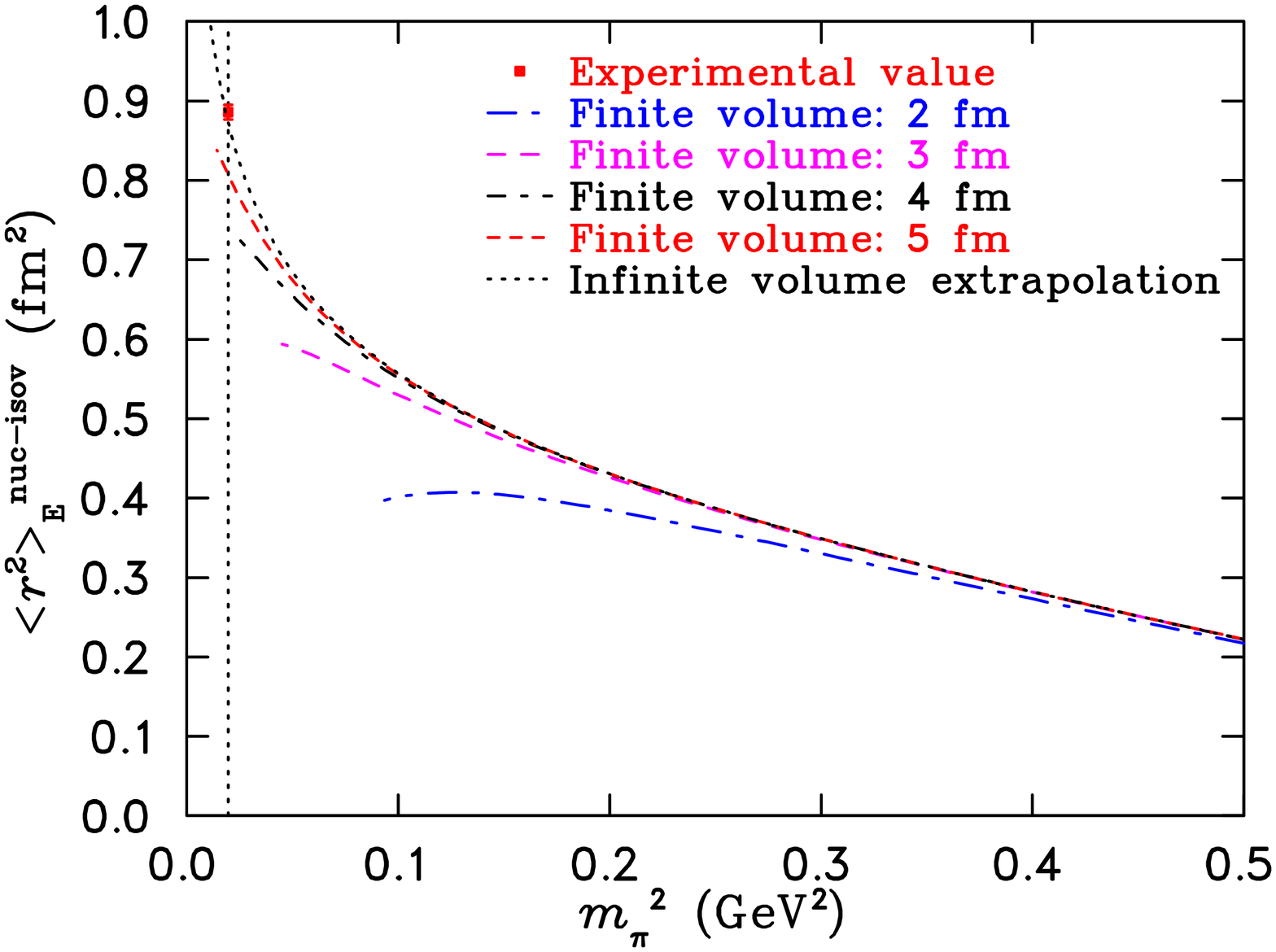}
\vspace{-11pt}
\caption{\footnotesize{(color online). Extrapolations of $\rad_E$ 
at different finite volumes, and at infinite volume. 
The curves are based on lattice QCD results from QCDSF, 
lattice sizes: $1.7-2.9$ fm, and the experimental value. 
The provisional constraint $m_\pi L > 3$ 
is used. The experimental value \cite{Mohr:2012tt,Beringer:1900zz} 
is marked as a square. 
}}
\label{fig:extrapbare}
\vspace{3mm}
\includegraphics[height=0.50\hsize,angle=90]{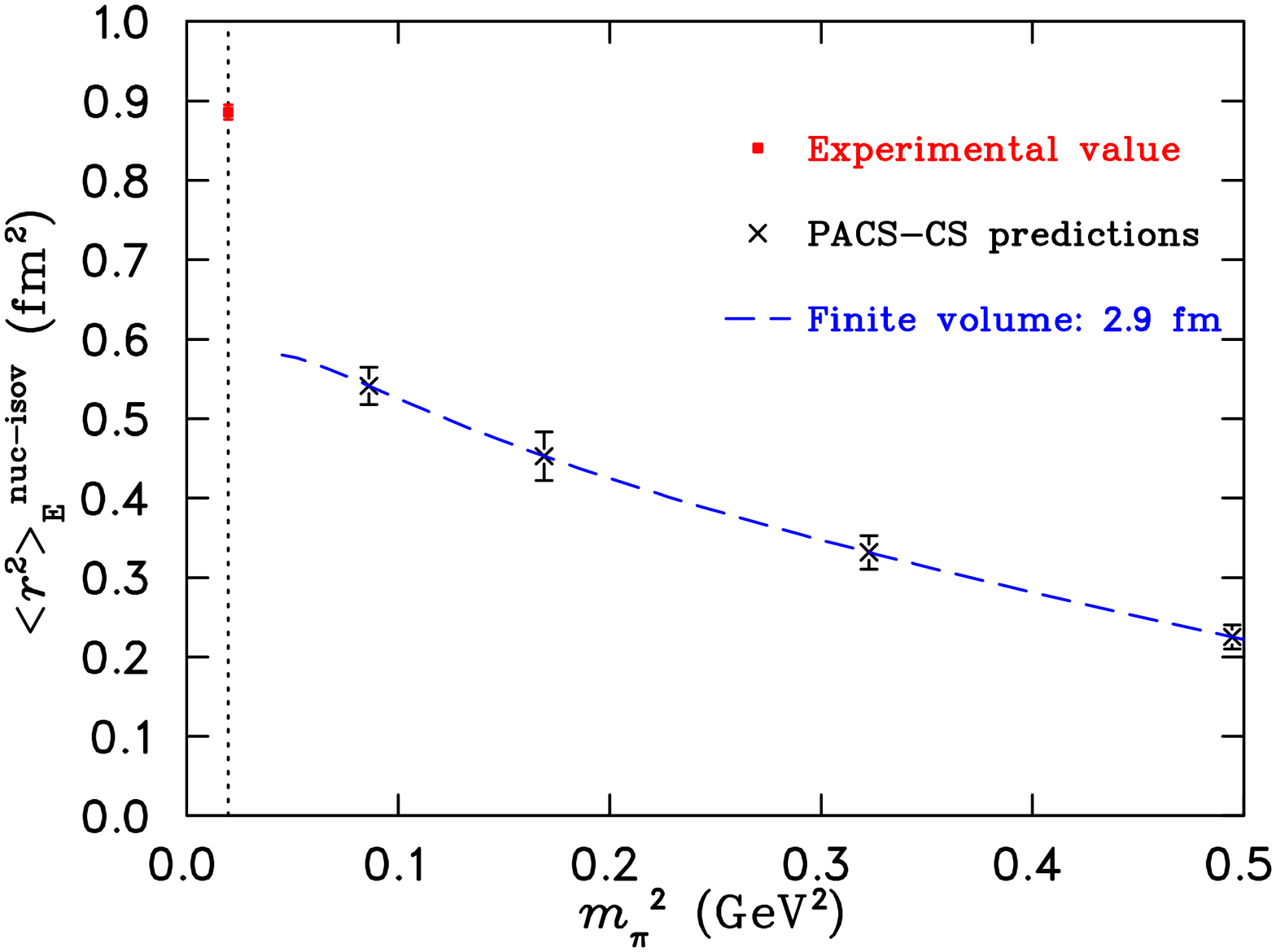}
\vspace{-11pt}
\caption{\footnotesize{(color online). Predictions of $\rad_E$ 
based on the volume ($L = 2.9$ fm) 
and pion masses from the PACS-CS lattice QCD configurations 
\cite{Aoki:2008sm}. The error bars represent the total uncertainties. The 
points are estimated only within the constraint $m_\pi L > 3$. 
}}
\label{fig:extrapPACSCS}
\end{figure*}

To predict the finite-volume dependence of future lattice simulations, 
consider again 
the electric charge radii from the lattice, corrected to infinite volume 
obtained using the variant \textit{Ansatz}  
defined in Eq.~(\ref{eqn:quad}). As shown in Fig.~\ref{fig:extrapinf}, 
this time 
the experimental value is included in the fit, 
and the chiral extrapolation at infinite volume is shown. 

With the fit parameters determined, 
extrapolations at a variety of finite volumes 
are shown in Fig.~\ref{fig:extrapbare}. 
The extrapolations use the lightest seven data points, and 
 are only calculated for values $m_\pi L > 3$, 
as in the initial 
selection of the lattice simulation results. 
These finite-volume results 
allow comparisons  with current lattice simulations, 
 and also 
allow estimates of finite-volume effects at arbitrary box sizes to be made. 
For example, 
using a box size of $L\sim 4$ fm, a significant deviation 
from the infinite-volume limit is observed. 
In this case, the finite-volume radius is $\rad_E^{\ro{nuc-isov}} = 0.745$ fm; 
significantly below the physical value of $0.886$ fm 
used to constrain the fit. 

In addition, the finite-volume extrapolations can provide 
a benchmark for lattice QCD simulations at large and currently 
untested box sizes. 
The extrapolation
 curves indicate that a box length of $L\gtrsim 7\,\,{\rm fm}$  is required 
to achieve an extrapolation within $2\%$ of the infinite-volume 
result.

This extrapolation method may be used to provide specific predictions for
 the charge radius based on lattice configurations from the 
PACS-CS Collaboration \cite{Aoki:2008sm}, freely available via the 
International Lattice Data Grid (ILDG). 
By choosing the lattice volume 
and the $m_\pi^2$ values to match the PACS-CS data, 
an estimate of the expected charge radii to be observed in future 
lattice simulations are shown in 
Fig.~\ref{fig:extrapPACSCS}. 
It is noteworthy that the predicted values of the charge radius near 
the physical point do not approach the experimental 
point at the PACS-CS lattice volume of $L = 2.9$ fm. 
This emphasizes the importance of using $\chi$EFT to correct 
for finite-volume effects, until very large lattice volumes can be used to 
resolve the correct chiral nonanalytic behaviour of 
hadrons. 

\vspace{-2mm}
\section{Conclusion}
\label{sect:conc}
\vspace{-2mm}

Newly developed techniques within the framework of chiral effective 
field theory were applied to recent 
precision lattice QCD results from the QCDSF Collaboration for the 
charge radius of the isovector nucleon. 
The inclusion of chiral loop contributions is vital for reconciling 
lattice simulations with the experimental result. 
It was discovered that the logarithmic divergence in the 
chiral expansion of the charge radius 
drives the large finite-volume corrections encountered near the physical point. 
 Lattice box sizes of $L\gtrsim 7\,{\rm fm}$
 are required in order to achieve a direct lattice simulation 
result within $2\%$ of the value at the physical point.

A discrepancy was found between the experimental value 
and the extrapolation results, which may be 
a consequence of excited state contamination; the use of only two flavours,  
and/or neglecting $\ca{O}(a)$ contributions.  

Finite-volume chiral extrapolations provide a benchmark for 
future lattice simulations. Specific predictions can be made 
 by choosing lattice volumes and pion masses to match those of 
a lattice calculation. By using this method, estimates of the 
electric charge radii simulations 
were obtained based on the PACS-CS configurations, which 
 provide a guide for the interpretation of future lattice results.


\vspace{-4mm}
\begin{acknowledgments}
\vspace{-2mm}
We would like the thank James Zanotti for many helpful discussions. 
This research 
is supported by the Australian Research Council through Grants No. DP110101265 
and No. FT120100821 (R.D.Y.). 
\end{acknowledgments}

\bibliographystyle{apsrev}
\bibliography{elecref}

\begin{thebibliography}{39}
\expandafter\ifx\csname natexlab\endcsname\relax\def\natexlab#1{#1}\fi
\expandafter\ifx\csname bibnamefont\endcsname\relax
  \def\bibnamefont#1{#1}\fi
\expandafter\ifx\csname bibfnamefont\endcsname\relax
  \def\bibfnamefont#1{#1}\fi
\expandafter\ifx\csname citenamefont\endcsname\relax
  \def\citenamefont#1{#1}\fi
\expandafter\ifx\csname url\endcsname\relax
  \def\url#1{\texttt{#1}}\fi
\expandafter\ifx\csname urlprefix\endcsname\relax\def\urlprefix{URL }\fi
\providecommand{\bibinfo}[2]{#2}
\providecommand{\eprint}[2][]{\url{#2}}

\bibitem[{\citenamefont{Gao}(2003)}]{Gao:2003ag}
\bibinfo{author}{\bibfnamefont{H.-y.} \bibnamefont{Gao}},
  \bibinfo{journal}{Int.J.Mod.Phys.} \textbf{\bibinfo{volume}{E12}},
  \bibinfo{pages}{1} (\bibinfo{year}{2003}), \eprint{nucl-ex/0301002}.

\bibitem[{\citenamefont{Hyde and de~Jager}(2004)}]{HydeWright:2004gh}
\bibinfo{author}{\bibfnamefont{C.~E.} \bibnamefont{Hyde}} \bibnamefont{and}
  \bibinfo{author}{\bibfnamefont{K.}~\bibnamefont{de~Jager}},
  \bibinfo{journal}{Ann.Rev.Nucl.Part.Sci.} \textbf{\bibinfo{volume}{54}},
  \bibinfo{pages}{217} (\bibinfo{year}{2004}), \eprint{nucl-ex/0507001}.

\bibitem[{\citenamefont{Arrington et~al.}(2007)\citenamefont{Arrington,
  Roberts, and Zanotti}}]{Arrington:2006zm}
\bibinfo{author}{\bibfnamefont{J.}~\bibnamefont{Arrington}},
  \bibinfo{author}{\bibfnamefont{C.}~\bibnamefont{Roberts}}, \bibnamefont{and}
  \bibinfo{author}{\bibfnamefont{J.}~\bibnamefont{Zanotti}},
  \bibinfo{journal}{J.Phys.G} \textbf{\bibinfo{volume}{G34}},
  \bibinfo{pages}{S23} (\bibinfo{year}{2007}), \eprint{nucl-th/0611050}.

\bibitem[{\citenamefont{Perdrisat et~al.}(2007)\citenamefont{Perdrisat,
  Punjabi, and Vanderhaeghen}}]{Perdrisat:2006hj}
\bibinfo{author}{\bibfnamefont{C.}~\bibnamefont{Perdrisat}},
  \bibinfo{author}{\bibfnamefont{V.}~\bibnamefont{Punjabi}}, \bibnamefont{and}
  \bibinfo{author}{\bibfnamefont{M.}~\bibnamefont{Vanderhaeghen}},
  \bibinfo{journal}{Prog.Part.Nucl.Phys.} \textbf{\bibinfo{volume}{59}},
  \bibinfo{pages}{694} (\bibinfo{year}{2007}), \eprint{hep-ph/0612014}.

\bibitem[{\citenamefont{Arrington et~al.}(2011)\citenamefont{Arrington,
  de~Jager, and Perdrisat}}]{Arrington:2011kb}
\bibinfo{author}{\bibfnamefont{J.}~\bibnamefont{Arrington}},
  \bibinfo{author}{\bibfnamefont{K.}~\bibnamefont{de~Jager}}, \bibnamefont{and}
  \bibinfo{author}{\bibfnamefont{C.~F.} \bibnamefont{Perdrisat}},
  \bibinfo{journal}{J.Phys.Conf.Ser.} \textbf{\bibinfo{volume}{299}},
  \bibinfo{pages}{012002} (\bibinfo{year}{2011}), \eprint{1102.2463}.

\bibitem[{\citenamefont{Yamazaki et~al.}(2009)\citenamefont{Yamazaki, Aoki,
  Blum, Lin, Ohta et~al.}}]{Yamazaki:2009zq}
\bibinfo{author}{\bibfnamefont{T.}~\bibnamefont{Yamazaki}},
  \bibinfo{author}{\bibfnamefont{Y.}~\bibnamefont{Aoki}},
  \bibinfo{author}{\bibfnamefont{T.}~\bibnamefont{Blum}},
  \bibinfo{author}{\bibfnamefont{H.-W.} \bibnamefont{Lin}},
  \bibinfo{author}{\bibfnamefont{S.}~\bibnamefont{Ohta}}, \bibnamefont{et~al.},
  \bibinfo{journal}{Phys.Rev.} \textbf{\bibinfo{volume}{D79}},
  \bibinfo{pages}{114505} (\bibinfo{year}{2009}), \eprint{0904.2039}.

\bibitem[{\citenamefont{Syritsyn et~al.}(2010)\citenamefont{Syritsyn, Bratt,
  Lin, Meyer, Negele et~al.}}]{Syritsyn:2009mx}
\bibinfo{author}{\bibfnamefont{S.}~\bibnamefont{Syritsyn}},
  \bibinfo{author}{\bibfnamefont{J.}~\bibnamefont{Bratt}},
  \bibinfo{author}{\bibfnamefont{M.}~\bibnamefont{Lin}},
  \bibinfo{author}{\bibfnamefont{H.}~\bibnamefont{Meyer}},
  \bibinfo{author}{\bibfnamefont{J.}~\bibnamefont{Negele}},
  \bibnamefont{et~al.}, \bibinfo{journal}{Phys.Rev.}
  \textbf{\bibinfo{volume}{D81}}, \bibinfo{pages}{034507}
  (\bibinfo{year}{2010}), \eprint{0907.4194}.

\bibitem[{\citenamefont{Bratt et~al.}(2010)}]{Bratt:2010jn}
\bibinfo{author}{\bibfnamefont{J.}~\bibnamefont{Bratt}} \bibnamefont{et~al.}
  (\bibinfo{collaboration}{LHPC Collaboration}), \bibinfo{journal}{Phys.Rev.}
  \textbf{\bibinfo{volume}{D82}}, \bibinfo{pages}{094502}
  (\bibinfo{year}{2010}), \eprint{1001.3620}.

\bibitem[{\citenamefont{Alexandrou et~al.}(2011)\citenamefont{Alexandrou,
  Brinet, Carbonell, Constantinou, Harraud et~al.}}]{Alexandrou:2011db}
\bibinfo{author}{\bibfnamefont{C.}~\bibnamefont{Alexandrou}},
  \bibinfo{author}{\bibfnamefont{M.}~\bibnamefont{Brinet}},
  \bibinfo{author}{\bibfnamefont{J.}~\bibnamefont{Carbonell}},
  \bibinfo{author}{\bibfnamefont{M.}~\bibnamefont{Constantinou}},
  \bibinfo{author}{\bibfnamefont{P.}~\bibnamefont{Harraud}},
  \bibnamefont{et~al.}, \bibinfo{journal}{Phys.Rev.}
  \textbf{\bibinfo{volume}{D83}}, \bibinfo{pages}{094502}
  (\bibinfo{year}{2011}), \eprint{1102.2208}.

\bibitem[{\citenamefont{Collins et~al.}(2011)\citenamefont{Collins, Gockeler,
  Hagler, Horsley, Nakamura et~al.}}]{Collins:2011mk}
\bibinfo{author}{\bibfnamefont{S.}~\bibnamefont{Collins}},
  \bibinfo{author}{\bibfnamefont{M.}~\bibnamefont{Gockeler}},
  \bibinfo{author}{\bibfnamefont{P.}~\bibnamefont{Hagler}},
  \bibinfo{author}{\bibfnamefont{R.}~\bibnamefont{Horsley}},
  \bibinfo{author}{\bibfnamefont{Y.}~\bibnamefont{Nakamura}},
  \bibnamefont{et~al.}, \bibinfo{journal}{Phys.Rev.}
  \textbf{\bibinfo{volume}{D84}}, \bibinfo{pages}{074507}
  (\bibinfo{year}{2011}), \eprint{1106.3580}.

\bibitem[{\citenamefont{Tiburzi}(2008)}]{Tiburzi:2007ep}
\bibinfo{author}{\bibfnamefont{B.~C.} \bibnamefont{Tiburzi}},
  \bibinfo{journal}{Phys.Rev.} \textbf{\bibinfo{volume}{D77}},
  \bibinfo{pages}{014510} (\bibinfo{year}{2008}), \eprint{0710.3577}.

\bibitem[{\citenamefont{Hu et~al.}(2007)\citenamefont{Hu, Jiang, and
  Tiburzi}}]{Hu:2007eb}
\bibinfo{author}{\bibfnamefont{J.}~\bibnamefont{Hu}},
  \bibinfo{author}{\bibfnamefont{F.-J.} \bibnamefont{Jiang}}, \bibnamefont{and}
  \bibinfo{author}{\bibfnamefont{B.~C.} \bibnamefont{Tiburzi}},
  \bibinfo{journal}{Phys.Lett.} \textbf{\bibinfo{volume}{B653}},
  \bibinfo{pages}{350} (\bibinfo{year}{2007}), \eprint{0706.3408}.

\bibitem[{\citenamefont{Jiang and Tiburzi}(2008)}]{Jiang:2008ja}
\bibinfo{author}{\bibfnamefont{F.-J.} \bibnamefont{Jiang}} \bibnamefont{and}
  \bibinfo{author}{\bibfnamefont{B.}~\bibnamefont{Tiburzi}},
  \bibinfo{journal}{Phys.Rev.} \textbf{\bibinfo{volume}{D78}},
  \bibinfo{pages}{114505} (\bibinfo{year}{2008}), \eprint{0810.1495}.

\bibitem[{\citenamefont{Hall et~al.}(2012{\natexlab{a}})\citenamefont{Hall,
  Leinweber, Owen, and Young}}]{Hall:2012yx}
\bibinfo{author}{\bibfnamefont{J.}~\bibnamefont{Hall}},
  \bibinfo{author}{\bibfnamefont{D.}~\bibnamefont{Leinweber}},
  \bibinfo{author}{\bibfnamefont{B.}~\bibnamefont{Owen}}, \bibnamefont{and}
  \bibinfo{author}{\bibfnamefont{R.}~\bibnamefont{Young}}
  (\bibinfo{year}{2012}{\natexlab{a}}), \eprint{1210.6124}.

\bibitem[{\citenamefont{Leinweber et~al.}(2004)\citenamefont{Leinweber, Thomas,
  and Young}}]{Leinweber:2003dg}
\bibinfo{author}{\bibfnamefont{D.~B.} \bibnamefont{Leinweber}},
  \bibinfo{author}{\bibfnamefont{A.~W.} \bibnamefont{Thomas}},
  \bibnamefont{and} \bibinfo{author}{\bibfnamefont{R.~D.} \bibnamefont{Young}},
  \bibinfo{journal}{Phys.Rev.Lett.} \textbf{\bibinfo{volume}{92}},
  \bibinfo{pages}{242002} (\bibinfo{year}{2004}), \eprint{hep-lat/0302020}.

\bibitem[{\citenamefont{Leinweber et~al.}(2006)\citenamefont{Leinweber, Thomas,
  and Young}}]{Leinweber:2005cm}
\bibinfo{author}{\bibfnamefont{D.~B.} \bibnamefont{Leinweber}},
  \bibinfo{author}{\bibfnamefont{A.~W.} \bibnamefont{Thomas}},
  \bibnamefont{and} \bibinfo{author}{\bibfnamefont{R.~D.} \bibnamefont{Young}},
  \bibinfo{journal}{PoS} \textbf{\bibinfo{volume}{LAT2005}},
  \bibinfo{pages}{048} (\bibinfo{year}{2006}), \eprint{hep-lat/0510070}.

\bibitem[{\citenamefont{Leinweber et~al.}(2005)\citenamefont{Leinweber, Thomas,
  and Young}}]{Leinweber:2005xz}
\bibinfo{author}{\bibfnamefont{D.~B.} \bibnamefont{Leinweber}},
  \bibinfo{author}{\bibfnamefont{A.~W.} \bibnamefont{Thomas}},
  \bibnamefont{and} \bibinfo{author}{\bibfnamefont{R.~D.} \bibnamefont{Young}},
  \bibinfo{journal}{Nucl. Phys.} \textbf{\bibinfo{volume}{A755}},
  \bibinfo{pages}{59} (\bibinfo{year}{2005}), \eprint{hep-lat/0501028}.

\bibitem[{\citenamefont{Hall et~al.}(2010)\citenamefont{Hall, Leinweber, and
  Young}}]{Hall:2010ai}
\bibinfo{author}{\bibfnamefont{J.~M.~M.} \bibnamefont{Hall}},
  \bibinfo{author}{\bibfnamefont{D.~B.} \bibnamefont{Leinweber}},
  \bibnamefont{and} \bibinfo{author}{\bibfnamefont{R.~D.} \bibnamefont{Young}},
  \bibinfo{journal}{Phys. Rev.} \textbf{\bibinfo{volume}{D82}},
  \bibinfo{pages}{034010} (\bibinfo{year}{2010}), \eprint{1002.4924}.

\bibitem[{\citenamefont{Hall et~al.}(2011)\citenamefont{Hall, Lee, Leinweber,
  Liu, Mathur et~al.}}]{Hall:2011en}
\bibinfo{author}{\bibfnamefont{J.}~\bibnamefont{Hall}},
  \bibinfo{author}{\bibfnamefont{F.}~\bibnamefont{Lee}},
  \bibinfo{author}{\bibfnamefont{D.}~\bibnamefont{Leinweber}},
  \bibinfo{author}{\bibfnamefont{K.}~\bibnamefont{Liu}},
  \bibinfo{author}{\bibfnamefont{N.}~\bibnamefont{Mathur}},
  \bibnamefont{et~al.}, \bibinfo{journal}{Phys.Rev.}
  \textbf{\bibinfo{volume}{D84}}, \bibinfo{pages}{114011}
  (\bibinfo{year}{2011}), \eprint{1101.4411}.

\bibitem[{\citenamefont{Hall et~al.}(2012{\natexlab{b}})\citenamefont{Hall,
  Leinweber, and Young}}]{Hall:2012pk}
\bibinfo{author}{\bibfnamefont{J.}~\bibnamefont{Hall}},
  \bibinfo{author}{\bibfnamefont{D.}~\bibnamefont{Leinweber}},
  \bibnamefont{and} \bibinfo{author}{\bibfnamefont{R.}~\bibnamefont{Young}},
  \bibinfo{journal}{Phys.Rev.} \textbf{\bibinfo{volume}{D85}},
  \bibinfo{pages}{094502} (\bibinfo{year}{2012}{\natexlab{b}}),
  \eprint{1201.6114}.

\bibitem[{\citenamefont{Aubin et~al.}(2004)\citenamefont{Aubin, Bernard, DeTar,
  Osborn, Gottlieb et~al.}}]{Aubin:2004wf}
\bibinfo{author}{\bibfnamefont{C.}~\bibnamefont{Aubin}},
  \bibinfo{author}{\bibfnamefont{C.}~\bibnamefont{Bernard}},
  \bibinfo{author}{\bibfnamefont{C.}~\bibnamefont{DeTar}},
  \bibinfo{author}{\bibfnamefont{J.}~\bibnamefont{Osborn}},
  \bibinfo{author}{\bibfnamefont{S.}~\bibnamefont{Gottlieb}},
  \bibnamefont{et~al.}, \bibinfo{journal}{Phys.Rev.}
  \textbf{\bibinfo{volume}{D70}}, \bibinfo{pages}{094505}
  (\bibinfo{year}{2004}), \eprint{hep-lat/0402030}.

\bibitem[{\citenamefont{Mohr et~al.}(2012)\citenamefont{Mohr, Taylor, and
  Newell}}]{Mohr:2012tt}
\bibinfo{author}{\bibfnamefont{P.~J.} \bibnamefont{Mohr}},
  \bibinfo{author}{\bibfnamefont{B.~N.} \bibnamefont{Taylor}},
  \bibnamefont{and} \bibinfo{author}{\bibfnamefont{D.~B.}
  \bibnamefont{Newell}}, \bibinfo{journal}{Rev.Mod.Phys.}
  \textbf{\bibinfo{volume}{84}}, \bibinfo{pages}{1527} (\bibinfo{year}{2012}),
  \eprint{1203.5425}.

\bibitem[{\citenamefont{Beringer et~al.}(2012)}]{Beringer:1900zz}
\bibinfo{author}{\bibfnamefont{J.}~\bibnamefont{Beringer}} \bibnamefont{et~al.}
  (\bibinfo{collaboration}{Particle Data Group}), \bibinfo{journal}{Phys.Rev.}
  \textbf{\bibinfo{volume}{D86}}, \bibinfo{pages}{010001}
  (\bibinfo{year}{2012}).

\bibitem[{\citenamefont{Jenkins and
  Manohar}(1991{\natexlab{a}})}]{Jenkins:1990jv}
\bibinfo{author}{\bibfnamefont{E.~E.} \bibnamefont{Jenkins}} \bibnamefont{and}
  \bibinfo{author}{\bibfnamefont{A.~V.} \bibnamefont{Manohar}},
  \bibinfo{journal}{Phys. Lett.} \textbf{\bibinfo{volume}{B255}},
  \bibinfo{pages}{558} (\bibinfo{year}{1991}{\natexlab{a}}).

\bibitem[{\citenamefont{Jenkins}(1992)}]{Jenkins:1991ts}
\bibinfo{author}{\bibfnamefont{E.~E.} \bibnamefont{Jenkins}},
  \bibinfo{journal}{Nucl. Phys.} \textbf{\bibinfo{volume}{B368}},
  \bibinfo{pages}{190} (\bibinfo{year}{1992}).

\bibitem[{\citenamefont{Jenkins and
  Manohar}(1991{\natexlab{b}})}]{Jenkins:1991ne}
\bibinfo{author}{\bibfnamefont{E.~E.} \bibnamefont{Jenkins}} \bibnamefont{and}
  \bibinfo{author}{\bibfnamefont{A.~V.} \bibnamefont{Manohar}}
  (\bibinfo{year}{1991}{\natexlab{b}}), \bibinfo{note}{talk presented at the
  Workshop on Effective Field Theories of the Standard Model, Dobogoko,
  Hungary, Aug 1991}.

\bibitem[{\citenamefont{Labrenz and Sharpe}(1996)}]{Labrenz:1996jy}
\bibinfo{author}{\bibfnamefont{J.~N.} \bibnamefont{Labrenz}} \bibnamefont{and}
  \bibinfo{author}{\bibfnamefont{S.~R.} \bibnamefont{Sharpe}},
  \bibinfo{journal}{Phys. Rev.} \textbf{\bibinfo{volume}{D54}},
  \bibinfo{pages}{4595} (\bibinfo{year}{1996}), \eprint{hep-lat/9605034}.

\bibitem[{\citenamefont{Walker-Loud}(2005)}]{WalkerLoud:2004hf}
\bibinfo{author}{\bibfnamefont{A.}~\bibnamefont{Walker-Loud}},
  \bibinfo{journal}{Nucl. Phys.} \textbf{\bibinfo{volume}{A747}},
  \bibinfo{pages}{476} (\bibinfo{year}{2005}), \eprint{hep-lat/0405007}.

\bibitem[{\citenamefont{Wang et~al.}(2007)\citenamefont{Wang, Leinweber,
  Thomas, and Young}}]{Wang:2007iw}
\bibinfo{author}{\bibfnamefont{P.}~\bibnamefont{Wang}},
  \bibinfo{author}{\bibfnamefont{D.~B.} \bibnamefont{Leinweber}},
  \bibinfo{author}{\bibfnamefont{A.~W.} \bibnamefont{Thomas}},
  \bibnamefont{and} \bibinfo{author}{\bibfnamefont{R.~D.} \bibnamefont{Young}},
  \bibinfo{journal}{Phys. Rev.} \textbf{\bibinfo{volume}{D75}},
  \bibinfo{pages}{073012} (\bibinfo{year}{2007}), \eprint{hep-ph/0701082}.

\bibitem[{\citenamefont{Lebed}(1995)}]{Lebed:1994ga}
\bibinfo{author}{\bibfnamefont{R.~F.} \bibnamefont{Lebed}},
  \bibinfo{journal}{Phys. Rev.} \textbf{\bibinfo{volume}{D51}},
  \bibinfo{pages}{5039} (\bibinfo{year}{1995}), \eprint{hep-ph/9411204}.

\bibitem[{\citenamefont{Armour et~al.}(2006)\citenamefont{Armour, Allton,
  Leinweber, Thomas, and Young}}]{Armour:2005mk}
\bibinfo{author}{\bibfnamefont{W.}~\bibnamefont{Armour}},
  \bibinfo{author}{\bibfnamefont{C.~R.} \bibnamefont{Allton}},
  \bibinfo{author}{\bibfnamefont{D.~B.} \bibnamefont{Leinweber}},
  \bibinfo{author}{\bibfnamefont{A.~W.} \bibnamefont{Thomas}},
  \bibnamefont{and} \bibinfo{author}{\bibfnamefont{R.~D.} \bibnamefont{Young}},
  \bibinfo{journal}{J. Phys.} \textbf{\bibinfo{volume}{G32}},
  \bibinfo{pages}{971} (\bibinfo{year}{2006}), \eprint{hep-lat/0510078}.

\bibitem[{\citenamefont{Beane}(2004)}]{Beane:2004tw}
\bibinfo{author}{\bibfnamefont{S.~R.} \bibnamefont{Beane}},
  \bibinfo{journal}{Phys. Rev.} \textbf{\bibinfo{volume}{D70}},
  \bibinfo{pages}{034507} (\bibinfo{year}{2004}), \eprint{hep-lat/0403015}.

\bibitem[{\citenamefont{Ali~Khan et~al.}(2004)}]{AliKhan:2003cu}
\bibinfo{author}{\bibfnamefont{A.}~\bibnamefont{Ali~Khan}} \bibnamefont{et~al.}
  (\bibinfo{collaboration}{QCDSF-UKQCD}), \bibinfo{journal}{Nucl. Phys.}
  \textbf{\bibinfo{volume}{B689}}, \bibinfo{pages}{175} (\bibinfo{year}{2004}),
  \eprint{hep-lat/0312030}.

\bibitem[{\citenamefont{Gell-Mann et~al.}(1968)\citenamefont{Gell-Mann, Oakes,
  and Renner}}]{GellMann:1968rz}
\bibinfo{author}{\bibfnamefont{M.}~\bibnamefont{Gell-Mann}},
  \bibinfo{author}{\bibfnamefont{R.~J.} \bibnamefont{Oakes}}, \bibnamefont{and}
  \bibinfo{author}{\bibfnamefont{B.}~\bibnamefont{Renner}},
  \bibinfo{journal}{Phys. Rev.} \textbf{\bibinfo{volume}{175}},
  \bibinfo{pages}{2195} (\bibinfo{year}{1968}).

\bibitem[{\citenamefont{Young et~al.}(2003)\citenamefont{Young, Leinweber, and
  Thomas}}]{Young:2002ib}
\bibinfo{author}{\bibfnamefont{R.~D.} \bibnamefont{Young}},
  \bibinfo{author}{\bibfnamefont{D.~B.} \bibnamefont{Leinweber}},
  \bibnamefont{and} \bibinfo{author}{\bibfnamefont{A.~W.}
  \bibnamefont{Thomas}}, \bibinfo{journal}{Prog.Part.Nucl.Phys.}
  \textbf{\bibinfo{volume}{50}}, \bibinfo{pages}{399} (\bibinfo{year}{2003}),
  \eprint{hep-lat/0212031}.

\bibitem[{\citenamefont{Kelly}(2004)}]{Kelly:2004hm}
\bibinfo{author}{\bibfnamefont{J.}~\bibnamefont{Kelly}},
  \bibinfo{journal}{Phys.Rev.} \textbf{\bibinfo{volume}{C70}},
  \bibinfo{pages}{068202} (\bibinfo{year}{2004}).

\bibitem[{\citenamefont{Leinweber and Cohen}(1993)}]{Leinweber:1992hj}
\bibinfo{author}{\bibfnamefont{D.~B.} \bibnamefont{Leinweber}}
  \bibnamefont{and} \bibinfo{author}{\bibfnamefont{T.~D.} \bibnamefont{Cohen}},
  \bibinfo{journal}{Phys.Rev.} \textbf{\bibinfo{volume}{D47}},
  \bibinfo{pages}{2147} (\bibinfo{year}{1993}), \eprint{hep-lat/9211058}.

\bibitem[{\citenamefont{Djukanovic et~al.}(2005)\citenamefont{Djukanovic,
  Schindler, Gegelia, and Scherer}}]{Djukanovic:2004px}
\bibinfo{author}{\bibfnamefont{D.}~\bibnamefont{Djukanovic}},
  \bibinfo{author}{\bibfnamefont{M.~R.} \bibnamefont{Schindler}},
  \bibinfo{author}{\bibfnamefont{J.}~\bibnamefont{Gegelia}}, \bibnamefont{and}
  \bibinfo{author}{\bibfnamefont{S.}~\bibnamefont{Scherer}},
  \bibinfo{journal}{Phys. Rev.} \textbf{\bibinfo{volume}{D72}},
  \bibinfo{pages}{045002} (\bibinfo{year}{2005}), \eprint{hep-ph/0407170}.

\bibitem[{\citenamefont{Aoki et~al.}(2009)}]{Aoki:2008sm}
\bibinfo{author}{\bibfnamefont{S.}~\bibnamefont{Aoki}} \bibnamefont{et~al.}
  (\bibinfo{collaboration}{PACS-CS Collaboration}),
  \bibinfo{journal}{Phys.Rev.} \textbf{\bibinfo{volume}{D79}},
  \bibinfo{pages}{034503} (\bibinfo{year}{2009}), \eprint{0807.1661}.

\end{thebibliography}

%


\end{document}